\documentclass[journal]{IEEEtran}

\usepackage[most]{tcolorbox}

\usepackage{booktabs}
\usepackage{subfig}
\usepackage{xcolor}
\usepackage{amsmath}
\usepackage{graphicx}
\usepackage{tabularx}
\usepackage[
separate-uncertainty = true,
multi-part-units = repeat
]{siunitx}
\usepackage{supertabular,booktabs}
\usepackage{tikz}
\usepackage{textcomp}
\usepackage{hyperref}
\usepackage{url}

\newcommand{\note}[1]{#1}
\newcommand{\notea}[1]{#1}

\newcommand\copyrighttext{%
	\footnotesize \textcopyright 2021 IEEE. Personal use of this material is permitted.
	Permission from IEEE must be obtained for all other uses, in any current or future
	media, including reprinting/republishing this material for advertising or promotional
	purposes, creating new collective works, for resale or redistribution to servers or
	lists, or reuse of any copyrighted component of this work in other works.
	DOI: \href{https://ieeexplore.ieee.org/abstract/document/9440768}{10.1109/TCYB.2021.3073210}}
\newcommand\copyrightnotice{%
	\begin{tikzpicture}[remember picture,overlay]
	\node[anchor=south,yshift=10pt] at (current page.south) {\fbox{\parbox{\dimexpr\textwidth-\fboxsep-\fboxrule\relax}{\copyrighttext}}};
	\end{tikzpicture}%
}
\begin{document}

%opening
\title{Motor-Imagery-Based Brain Computer Interface using Signal Derivation and Aggregation Functions}

\author{Javier Fumanal-Idocin, Yu-Kai Wang ~\IEEEmembership{Member,~IEEE}, Chin-Teng Lin,~\IEEEmembership{Fellow,~IEEE}, Javier Fernández, Jose Antonio Sanz,
	Humberto Bustince,~\IEEEmembership{Senior,~IEEE}% <-this % stops a space
	
%\author[pamplona]{J. Fumanal-Idocin}
%\ead{javier.fumanal@unavarra.es}

%\author[sidney]{Yu-Kai Wang}
%\ead{YuKai.Wang@uts.edu.au}
%\author[sidney]{Chin-Teng Lin}
%\ead{Chin-Teng.Lin@uts.edu.au}

%\author[pamplona]{Javier Fernandez}
%\ead{fcojavier.fernandez@unavarra.es}
%\author[pamplona,arabia]{H. Bustince}
%\ead{bustince@unavarra.es}

%\address[pamplona]{Public University of Navarra and Institute of Smart Cities, Campus Arrosadia s/n, 31006 Pamplona, Spain}
%\address[sidney]{Centre for Artificial Intelligence, CIBCI Lab, Faculty of Engineering and Information Technology,	University of Technology Sydney, Sydney, Australia}
%\address[arabia]{Faculty of Computing and Information Technology, King Abdulaziz University, Jeddah, Saudi Arabia}

\thanks{Javier Fumanal-Idocin, Javier Fernandez, Jose Antonio Sanz and Humberto Bustince are with the Departamento de Estadistica, Informatica y Matematicas, Universidad Publica de Navarra, Campus de Arrosadia, 31006, Pamplona, Spain.
	emails: javier.fumanal@unavarra.es, fcojavier.fernandez@unavarra.es , joseantonio.sanz@unavarra.es, bustince@unavarra.es}% <-this % stops a space
\thanks{Javier Fernandez, Jose Antonio Sanz and Humberto Bustince are with the Institute of Smart Cities, Universidad Publica de Navarra, Campus de Arrosadia, 31006, Pamplona, Spain.}
\thanks{Javier Fernandez and Humberto Bustince are with the Laboratory Navarrabiomed, Hospital Complex of Navarre (CHN), Universidad Publica de Navarra, IdiSNA, Irunlarrea 3. 31008, Pamplona, Spain.}
\thanks{Y.-K. Wang and C.-T. Lin are with the Australian AI Institute,
	Faculty of Engineering and Information Technology, University of Technology Sydney, Ultimo, NSW 2007, Australia (e-mail:, yukai.wang@uts.edu.au;
	chin-teng.lin@uts.edu.au)}

}
%\markboth{IEEE Transactions on Cybernetics}%

\maketitle
\copyrightnotice
\begin{abstract}

\note{Brain Computer Interface (BCI) technologies are popular methods of communication between the human brain and external devices. One of the most popular approaches to BCI is Motor Imagery (MI)}. In BCI applications, the ElectroEncephaloGraphy (EEG) is \note{a very popular} measurement for brain dynamics because of its non-invasive nature. Although there is a high interest in the BCI topic, the performance of existing systems is still far from ideal, due to the difficulty of performing pattern recognition tasks in EEG signals. \note{This difficulty lies in the selection of the correct EEG channels, the signal-to-noise ratio of these \note{signals} and how to discern the redundant information among them}. BCI systems are composed of a wide range of components that perform signal \note{pre-processing}, feature extraction and decision making. In this paper, we define a new BCI Framework,
named Enhanced Fusion Framework, where we propose three different ideas to improve the existing  MI-based BCI frameworks. Firstly, we include \notea{an additional} \note{pre-processing step of the signal: }a differentiation of the EEG signal that makes it time-invariant. Secondly, we add \notea{an additional frequency band as feature for the system: the Sensory Motor Rhythm band, and we show its effect on the performance of the system}. Finally, \note{we make a profound study of how to make the final decision in the system.} We propose the usage of both up to six types of different classifiers and a wide range of aggregation functions (including classical aggregations, Choquet and Sugeno integrals and their extensions and overlap functions) to fuse the information given by the considered classifiers. We have tested this new system on a dataset of 20 volunteers performing motor imagery-based brain-computer interface experiments. On this dataset, the new system achieved a $88.80\%$ of accuracy. We also propose an optimized version of our system that is able to obtain up to $90,76\%$. Furthermore, we find that the pair Choquet/Sugeno integrals and
overlap functions are the ones providing the best results.

\end{abstract}


\section{Introduction}
Brain-computer interfaces (BCIs) provide a new interface between the human brain and the devices or systems to be controlled by the changes of brain dynamics \cite{lin2018wireless}, \cite{wang2015eeg}. One popular BCI is Motor Imagery (MI) in which a person imagines a specific body movement (usually moving left or right hand). As \note{imagining} the movement, the Event-Related Desynchronization (ERD) in mu rhythm near motor areas has been widely reported in the previous studies \cite{ko2019multimodal}, \cite{lance2015towards}. Therefore, correct ERD identification highly influences the performance of MI-based BCI. Recently, a lot of state-of-the-art algorithms, such as Common Spatial Pattern (CSP), support vector machines, or deep learning have been extensively used to identify the ERD in MI-based BCI \cite{lance2015towards, tabar2016novel, wang2015eeg, 8386437}.

BCIs use a wide range of techniques to extract features from the original raw data. Due to the volume-conduction effect, it is very difficult to extract information directly from the ElectroEncephaloGraphy (EEG) data \cite{blankertz2011single}, \note{because the measures taken are affected by the conductance of the biological tissues that transmit the electrical signal}. To cope with this, most algorithms use a procedure to extract features from the EEG data before feeding them to a classifier. Some of the most common procedures include using the Fast Fourier transform (FFt) to transform the EEG signals to the frequency domain \cite{7300244, zheng2019emotions, murugappan2014wireless} and the Meyer wavelet transformation \cite{iacovello2016stimuli, 8337847}. There have also been many different fuzzy approaches to the BCI problem \cite{iran2017tfs, nott2017tfs, multi2019tfs}.

A BCI framework is composed of \note{signal pre-processing}, feature extraction and control commands. The interactions among all of these elements take a crucial role in the final performance of the system. However, possible correlations and synergies among the different features are ignored in the command control phase in the classical BCI framework. \note{In \cite{JIN2019262} the authors proposed a channel selection procedure to minimize the number of correlated components in the system. In \cite{feng2018towards} the authors use a time-window selection algorithm to choose the best time to collect the MI features, and the authors in \cite{qi2020spatiotemporal} the authors use the spatio-temporal information of the EEG signals to detect the optimal channels to discriminate between the MI tasks.} In \cite{ko2019multimodal} the authors propose a new BCI framework, the Multimodal Fuzzy Fusion BCI Framework (MFF), that uses the fuzzy integrals \cite{sugeno1974theory, beliakov2016practical} to model these interactions, and a two-step decision making process that considers that combines the outputs of three classical classifiers.

%Fuzzy integrals have been widely used in many fields \cite{grabisch1996application, beliakov2016practical, grabisch2010fuzzy}. 
Two of the most important fuzzy integrals are the Sugeno and the Choquet integrals. Both aggregate values using a measure that indicates how important are the different correlations among the data. Therefore, they are specially suited in applications where there are significant interactions among the features to aggregate. Fuzzy integrals have been widely used in decision making \cite{grabisch2000application}, image processing \cite{beliakov2016practical} and deep learning \cite{dias2018using}. As we have mentioned, Fuzzy integrals have already been used in a BCI framework in \cite{ko2019multimodal}, obtaining better results than the classical aggregations. Many different generalizations of the Choquet integral have been proposed \cite{auephanwiriyakul2002generalized, murofushi1991fuzzy, dimuro2020state}. The CF, \cite{LUCCA201894}, and C$_{F1,F2}$ \cite{dimuro2020generalized} generalizations of the Choquet integral have proven to be very successful in classification systems. Ordered-Weighted-Averaging operators \cite{yager2012ordered, yager2004generalized} (OWAs) are a specific case of the fuzzy integrals that have also been used in multicriteria decision \cite{de2017algorithm}, and finance \cite{merigo2011uncertain}.

A closed concept to aggregation functions are the overlap functions, which were introduced in \cite{bustince2010overlap} in the fuzzy community, as a way to represent the overlapping between two concepts. Since these functions were only defined for two elements, the generalized version of overlap functions for n-valued vectors were proposed in \cite{DEMIGUEL201981}. Some common examples of these generalized overlap functions are the geometric and the harmonic means. Generalized overlap functions have been successfully used in Big Data \cite{elkano2016fuzzy} and in fuzzy rule-based classifiers \cite{elkano2014enhancing}. 

The most successful MI-based BCI framework using aggregation functions is  \cite{ko2019multimodal}. However, in the decision making process, it does not study:
\begin{enumerate}
	\item The effects of new types of classifiers with the new integrals.
	\item The possibility of using different aggregation functions in each step of the process.
	\item It does not  improve other areas of the BCI framework besides the decision making phase.
\end{enumerate}  

In this paper, we present a new BCI framework, named Enhanced Fusion BCI Framework (EMF). It includes a new \notea{differentiation signal phase, an additional wave band: the SensoriMotor Rhythm, and we add two additional types of classifiers to the ensemble of classifiers: Gaussian Process and Support Vector Machines. We also consider a wider set of aggregation functions to be used in the decision making phase that includes not only the Choquet and Sugeno integrals and their generalizations, but also Ordered Weighted Averaging operators and generalized overlaps}. Finally, we also propose an Optimized version
of the EMF (OEMF) in terms of accuracy by checking the most proper combinations of wave bands and classifiers.

The rest of our paper is organized as follows. In section \ref{sec:preliminars} we remind the concept of is an aggregation functions and different types of them, we also describe the traditional BCI framework \note{\cite{wu2016fuzzy}} and the MFF BCI framework \cite{ko2019multimodal}. In section \ref{sec:emf_bci} we explain the the new Enhanced Multimodal Fusion BCI Framework. In section \ref{sec:experiments} we show our experimental results for our own BCI dataset, and in section \ref{sec:bciiv_dataset} we discuss our results for the BCI IV competition dataset \cite{tangermann2012review}. Finally, in section \ref{sec:conclusions} we give our final conclusions and remarks for this work.

\section{Preliminars} \label{sec:preliminars}

In this section we recall some basic notions about aggregation functions (Section \ref{sec:agg}), the traditional BCI framework (Section \ref{sec:tradBCI}) and the MFF BCI framework (Section \ref{subsec:mmf_bci}).
\subsection{Aggregation Functions}
\label{sec:agg}

Aggregation functions are used to fuse information from n sources into one single output. A function A: $[0, 1]^n \rightarrow [0,1]$ is said to be a n-ary aggregation function if the following conditions hold:
\begin{itemize}
	\item A is increasing in each argument: $\forall i \in \{1,\dots,n\}, i<y, A(x_1, \dots.,x_i,\dots x_n) \le A(x_1, \dots,y,\dots x_n)$
	\item $A(0,\dots,0) = 0$
	\item $A(1,\dots,1) = 1$
\end{itemize}

Some examples of classical n-ary aggregation functions are: 
\begin{itemize}
	\item Artihmetic mean: $A(x) = \frac{1}{n}\sum_{i=1}^{n} x_i $.
	\item Median: $A(x) = x_i: \{a : \forall x_a < x_i\}, \{b, \forall x_b > x_i \}, |a| = |b|$.
	\item Max: $A(x) = max(x_1, \dots, x_n)$.
	\item Min: $A(x) = min(x_1, \dots, x_n)$.
\end{itemize}

%Where $x = (x_1, ..., x_n)$.

Other types of aggregation functions are the following ones:

%The whole list of aggregation functions tested in the EMF are:
%\begin{itemize}
%	\item Mean.
%	\item Median.
%	\item Minimum.
%	\item Maximum.
%	\item Choquet integral.
%	\item CF$_{min,min}$.
%	\item CF$_{\sqrt{x,y},T_{Luk}(x,y)}$.
%	\item Choquet-Hamacher.
%	\item Sugeno.
%	\item Sugeno-Hamacher.
%	\item OWA$_1$
%	\item OWA$_2$
%	\item OWA$_3$
%\end{itemize}

\subsubsection{T-norm \cite{GUPTA1991431}}
A T-norm is an aggregation function $[0, 1]^2 \rightarrow [0, 1]$that satisfies the following properties for $a,b,c \in [0,1]$:
\begin{itemize}
	\item $T(a,b) = T(b, a)$
	\item $T(a, T(b, c)) = T(T(a,b), c)$
	\item $T(a, 1) = a$
\end{itemize}
Some examples of T-norms are the product or the minimum.
\subsubsection{Choquet integral \cite{beliakov2016practical}}
Having $N=\{1,\dots, n\}$, a function $m: 2^n \rightarrow [0,1]$ is a fuzzy measure if, for all $X,Y \in N$, it satisfies the following properties: 
\begin{enumerate}
\item[($m$1)] Increasingness: if $X\in Y$, then $m(X) \subseteq m(Y)$.
\item[($m$2)] Boundary conditions: $m(\emptyset) =0$ and $m(N)=1$.

\end{enumerate}

The discrete Choquet integral of $\textbf{x} = (x_1,\dots,x_n)\in [0,1]^n$ with respect to $m$ is defined as $C_m: [0,1]^n\rightarrow [0,1]$ given by

\begin{equation} \label{eq:choquet}
	C_m(x) = \sum_{i=1}^{n} (x_{\sigma(i)}-x_{\sigma(i-1)}) \cdot m(A_i)
\end{equation}

where $\textbf{x}_\sigma$ is an increasing permutation of \textbf{x} such that $0\le x_{\sigma(1)} \le \dots \le x_{\sigma(n)}$. With the convention that $x_0 = 0$, and $A_i=\{i,\dots,n\}$.

\subsubsection{CF \cite{LUCCA201894}}
It is a generalization of the Choquet integral that replaces the product used in Eq. \ref{eq:choquet}  for a more general function F. In \cite{tfspreagregaciones} the authors detail the required properties for F so that the CF is an aggregation function, \note{and conclude that the best F in their experimental results is the Hamacher T-norm. For this reason, we have chosen it for our experimentation}, as detailed in the following expressions:

\[
T_H(x, y) = 
\begin{cases} 
0, & \mbox{if } x=y=0\\ \frac{xy}{x+y-xy}, & \mbox{otherwise} 
\end{cases}
\]

\[
CF(x) = \sum_{i=1}^{n} T_H(x_{\sigma(i)}-x_{\sigma(i-1)}, m(A_i))
\]
\subsubsection{C$_{F1,F2}$ \cite{dimuro2020generalized}}
The original product of the Choquet Integral can be decomposed on two product functions using the distributive property of the product. Therefore, the Choquet integral can be written as:
\[
C(x) = \sum_{i=1}^{n} x_{\sigma(i)}m(A_i)- x_{\sigma(i-1)} m(A_i)
\]
Then, the product functions are substituted for two more generic functions: F1 and F2. In \cite{dimuro2020generalized} the authors explain the properties that must hold F1 and F2 so that the C$_{F1,F2}$ is an aggregation function. Consequently, the expression for the C$_{F1,F2}$ is the following:
\[
C_{F1,F2}(x) = \sum_{i=1}^{n} F_1(x_{\sigma(i)}), m(A_i))- F_2(x_{\sigma(i-1)}), m(A_i))
\]
\subsubsection{Sugeno integral \cite{sugeno1974theory}}
Let $m:2^N \rightarrow [0,1]$ be a fuzzy measure. The discrete Sugeno integral of $\textbf{x}=(x_1, \dots, x_n) \in [0,1]^n$ with respect to $m$ is defined as a function $S_m:[0,1]^n\rightarrow [0,1]$, given by:

\begin{equation} \label{eq:sugeno}
	S_m(x) = \max \{min(x_{\sigma(i)}, m(A_i)) | i=1,..,n\}
\end{equation}

\subsubsection{Sugeno Hamacher \cite{ko2019multimodal}}

If we consider using the Hamacher T-norm instead of the minimum in Eq. \ref{eq:sugeno}, we obtain the following expression:

\[
S(x) = \max \{T_H(x_{\sigma(i)}, m(A_i)) | i=1,..,n\}
\]

\subsubsection{Ordered Weighted Averaging operators (OWA) \cite{yager2012ordered}}
$\overrightarrow{w} = (w_1, ..., w_n) \in [0,1]^n$ is called a weighting vector if $\sum_{i=1}^{n} w_i = 1$. The OWA operator associated to $\overrightarrow{w}$ is the mapping OWA$_{\overrightarrow{w}}: [0, 1]^n \rightarrow [0,1]$ defined for every $(x_1, ..., x_n) \in [0,1]^n$ by:
%They use use a quantifier function, $Q$, to generate a set of weights, $w$, to compute the following expression:

\[
OWA(x_1,...,x_n) = w_1x_{\gamma(1)} + ... + w_nx_{\gamma(n)}
\]

where $\gamma : \{1,...,n\} \rightarrow \{1,..,n\}$ is a permutation such that:
$x_{\gamma(1)} \ge x_{\gamma(2)} \ge ... \ge x_{\gamma(n)}$.

%Each $w_i$ is computed as:
The weight vector can be computed used a quantifier function, Q. For this study, we have used the following one:
\[
w_i = Q(\frac{i}{n}) - Q(\frac{i-1}{n})
\]
\[
Q_{a,b}(i) = 
\begin{cases} 
0, & \mbox{if } i<a \\ 
1, & \mbox{if } i>b \\
\frac{i-a}{b-a}, & \mbox{otherwise} 
\end{cases}
\]
where $a,b \in [0,1]$. Depending on the value of the parameters $a$ and $b$, different weight vectors can be obtained. We have used three different ones:
\begin{itemize}
	\item OWA$_1$: $a=0.1, b=0.5$
	\item OWA$_2$: $a=0.5, b=1$
	\item OWA$_3$: $a=0.3, b=0.8$
\end{itemize}

\subsubsection{Overlap functions \cite{DEMIGUEL201981}}
A n-dimensional overlap, $G$, is a $[0, 1]^n \rightarrow [0, 1]$ function that holds:
\begin{itemize}
	\item Is commutative.
	\item If $\prod_{i=1} x_i = 0$, then $G(x)=0$.
	\item If $\prod_{i=1} x_i = 1$, then $G(x)=1$
	\item $G$ is increasing.
	\item $G$ is continuous.
\end{itemize}

The minimum function, for example, is an overlap function. We have also considered three more:

\begin{itemize}
	\setlength\itemsep{1em}
	\item Harmonic Mean (HM): $\frac{n}{\sum_{i=1}^n \frac{1}{x_i}}$
	\item Sinus Overlap (SO): $sin \frac{\pi}{2}(\prod_{i=1}^{n} x_i)$
	\item Geometrical Mean (GM): $\sqrt[n]{\prod x_i}$
\end{itemize} 

\subsection{Traditional BCI Framework} 
\label{sec:tradBCI}
The traditional BCI system structure includes four parts:

\begin{enumerate}
	\item The first step is acquiring the EEG data from the commercial EEG device and performing band-pass filtering	and artefact removal on the collected EEG signals. 
	
	\item The second
	step is EEG feature transformation and feature extraction. Usually, the FFt is used to transform
	the EEG signals from the  into different frequency components \cite{akin2002comparison}. The FFt
	analysis transforms the time-series EEG signals in each channel into \note{its constituent frequencies}. \note{Following the procedure in \cite{wu2016fuzzy, qi2020spatiotemporal}, we cover the frequencies range 1-30Hz}. \notea{We study for the delta ($\delta$) wave band the 1-3 Hz frequencies, for the theta ($\theta$) wave band the 4-7 Hz frequencies, for the alpha ($\alpha$) 8-13 Hz frequencies, for the beta ($\beta$) the 14-30 Hz frequencies and All 1-30Hz frequencies \cite{teplan2002fundamentals}} using a 50-point moving window segment overlapping 45 data points.
	
	\item Subsequently, the CSP was
	used for feature extraction to extract the maximum spatial separability from the different EEG signals corresponding to the control commands. The CSP is a well-known supervised mathematical procedure commonly used in EEG signal processing. The CSP is used to transform multivariate EEG signals into well-separated subcomponents with maximum spatial variation using the labels for each example \cite{blankertz2007optimizing}, \cite{guger2000real}, \cite{gramfort2013meg}.
	
	\item Last, pattern classification is performed on the
	CSP features signals using an ensemble of classifiers
	to differentiate the commands. Each base classifier is trained using a different wave band (for instance, if the base classifier is the LDA, the ensemble would be composed of: $\delta-LDA$, $\theta-LDA$, $\alpha-LDA$, $\beta-LDA$, and $All-LDA$) and the final decision is taken combining all of them. The most common way of obtaining the final decision is to compute the arithmetic mean of the outputs of all the base classifiers (each one provides a probability for each class), and take the class with higher aggregated value. The most common classifiers used for this task are the Linear Discriminant Analysis (LDA),
	Quadratic Discriminant Analysis (QDA) and k-nearest neighbours
	classifier (KNN) \cite{tibarewala2010performance}.
\end{enumerate}

\subsection{Multimodal Fuzzy Framework} \label{subsec:mmf_bci}

The Multi-modal Fuzzy Framework (MFF) is proposed in \cite{ko2019multimodal}. It follows a similar structure to the one in the traditional BCI framework: it starts with the EEG measurements, it computes the FFt transformation to the frequency domain and it uses the CSP transform to obtain a set of features to train the classifiers.

However, in the MFF it is necessary to train not one, but three classifiers for each wave band: a LDA, a QDA and a KNN. We name the classifiers according to their type of classifier and the wave band used to train it. For instance, for the $\delta$ band we would have $\delta-LDA$, $\delta-QDA$ and $\delta-KNN$.

Then, the decision making phase is performed in two phases:
\begin{enumerate}
	\item Frequency phase: since we got a LDA, QDA and KNN for each wave band, the first step is to fuse the outputs of these classifiers in each wave band. For example, in the case of the LDA classifiers, we have a $\delta-LDA$, $\theta-LDA$, $\alpha-LDA$, $\beta-LDA$ and $All-LDA$ that will be fused using an aggregation function to obtain a vector, $FF-LDA$. That is, the same process explained for the traditional framework is applied but without making the final decision. We do the same with the QDA and KNN classifiers. The result of this phase is a list of collective vectors (one for each type of classifier).
	\item Classifier phase: in this phase, the input is the list of collective vectors given by each different kind of classifier ($FF-LDA$, $FF-LDA$,$FF-KNN$) computed in the frequency phase. We fuse the three vectors according to the classes, and the result is a vector containing the score for each class for the given sample. As in the traditional framework, the decision is made in favour to the class associated with the largest value.
\end{enumerate}
We must point out that the same aggregation is used for both the frequency phase and the classifier phases.

The aggregation functions tested in the MFF are the Choquet integral, the CF integral using the Hamacher T-norm, the $CF_{min,min}$ generalizations, the Sugeno integral and the Hamacher Sugeno integral. \note{We used the cardinal fuzzy measure for all of them \cite{8858815}}  

\section{Enhanced Fusion Framework} \label{sec:emf_bci}

\begin{figure*}
	\tcbox[colback=white!30, title=Enhanced Fusion Framework]{\includegraphics[width=\linewidth]{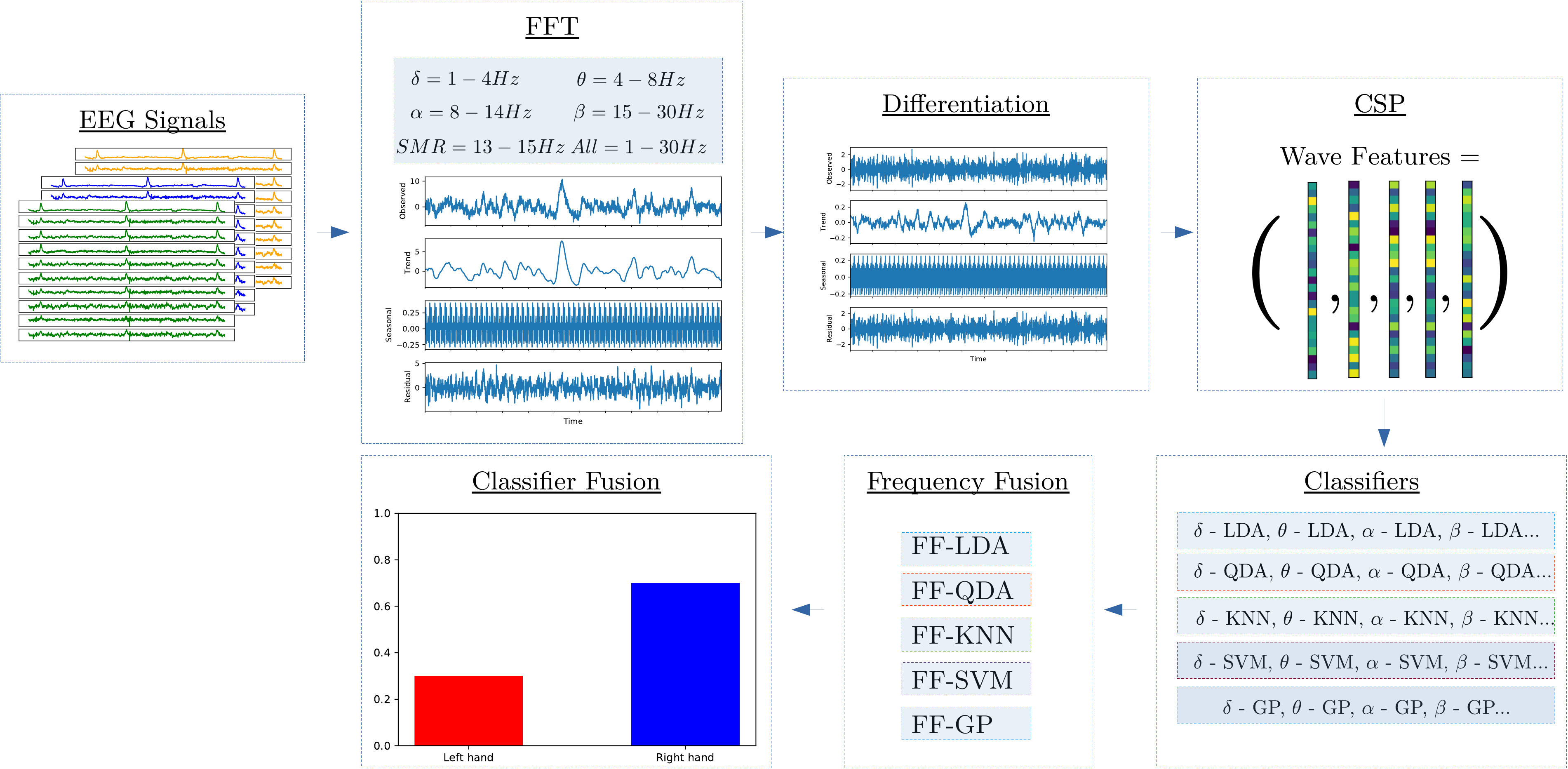}}
	\caption{Architecture of the proposed Enhanced Fusion Framework. }
	\label{fig:flow}
\end{figure*}

Our aim with the Enhanced Multimodal Fusion Framework (EMF) is to build upon the foundations of the MFF in order to improve its experimental results. Starting from the MFF, we add a new band as well as a new \note{signal pre-processing} phase known as differentiation. Furthermore, we considered more classifiers and a wider set of aggregation functions for the decision making process. Finally, we give more flexibility to the decision making process because we allow the aggregation function to be different in each stage.

A schematic view of the whole EMF classification process is in Fig. \ref{fig:flow}. The new components compared to the MFF in \cite{ko2019multimodal} can be seen in the figure: the SMR in the FFt phase, the differentiation phase,  and the two new classifiers (SVM and GP). In the following sections we present in detail each component added in the EMF framework.

\subsection{Wave bands}

For the EMF we have considered all the wave bands used in the traditional BCI framework: $\delta$, $\theta$, $\alpha$, $\beta$ and $All$. However, we have also added the SensoriMotor Rhythm (SMR), which covers the $13-15$Hz frequencies \cite{arroyo1993functional}.

\note{Regarding the nature of ERD/ERS, movement or preparation for movement is typically accompanied by a decrease in mu and beta rhythms, particularly contralateral to the movement \cite{wolpaw2002brain} to control sequence. This phenomenon has been named event-related desynchronization or ERD \cite{babiloni1999human}. Its opposite, rhythm increase, or event-related synchronization (ERS), occurs after movement and with relaxation \cite{pfurtscheller1999event}. So, we suppose that the activity present in this band will be closely related to the studied task.}

\subsection{Signal pre-processing \note{and feature extraction}}
The data acquisition process is similar to the rest of the BCI frameworks. First, we obtain the EEG data, second, we apply the FFt. Then, we add a new component, the differentiation for the signals. Finally, we compute the CSP of 25 components to extract the features from which we will train the classifiers from the differenced signals.

\begin{figure}
	\centering
	\subfloat[width=0.45\linewidth][Original activity.]{\includegraphics[width=0.45\linewidth]{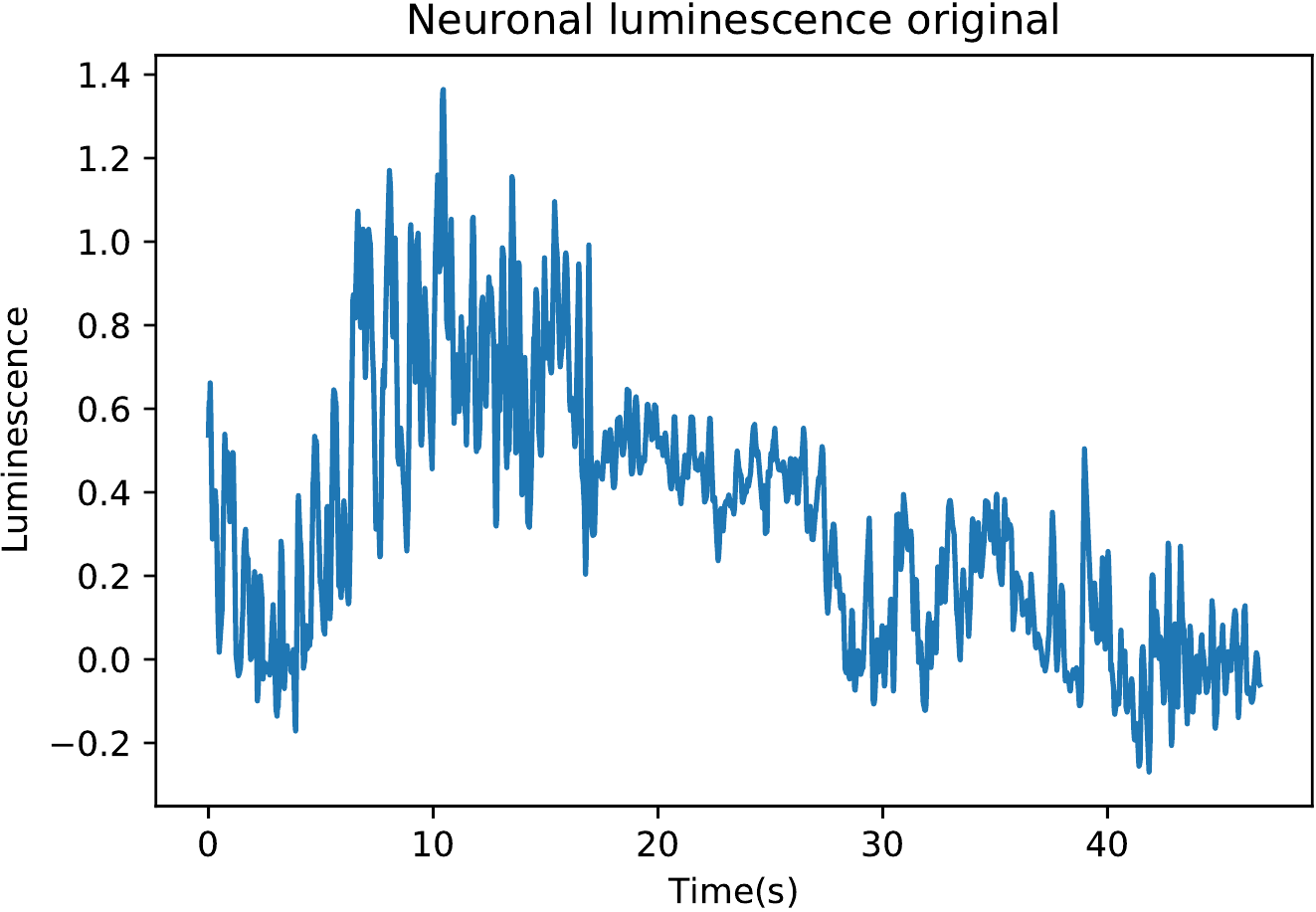}}
	\qquad
	\subfloat[width=0.45\linewidth][Filtered activity using the high pass filter \cite{aguilera2017signatures}.]{\includegraphics[width=0.45\linewidth]{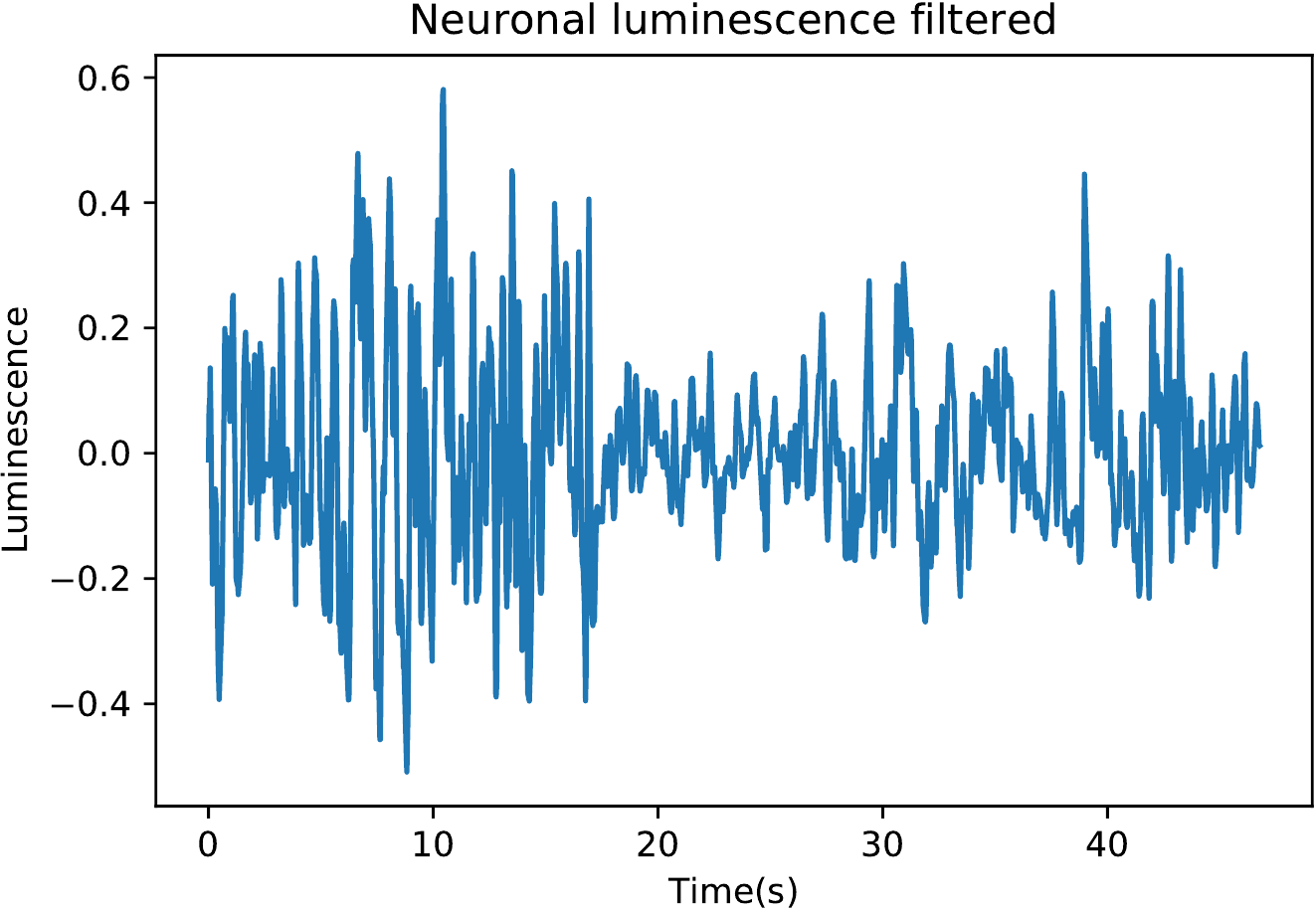}}
	\caption{Neuronal luminescence for a single neuron in a C.Elegans during free movement in one trial, \cite{nguyen2016whole}.}
	\label{fig:activityoriginal}
\end{figure}

 The idea to apply the differentiation comes from a related area, neuroscience. In \cite{nguyen2016whole} the authors measure the activations of a moving C.Elegans, using the luminescence of each neuron during a series of trials. Alongside the paper, the corresponding dataset was released. This dataset was composed of real numbers that quantified the luminescence of each neuron, instead of a neuron activated/deactivated classification, since a straightforward method to compute if a neuron was activated or not was not found. Authors in \cite{aguilera2017signatures}, based on this same C.Elegans dataset, stated that the real activations of the neurons should be computed not using the absolute value of the luminescence, but only the spikes of the signal. They attributed the big changes in the tendency of the temporal series to artefacts in the measures. So, they performed a low-pass filter to the signal (Fig. \ref{fig:activityoriginal}).

\begin{figure}
	\centering
%	\subfloat[width=0.45\linewidth][Left trials.]{\includegraphics[width=0.45\linewidth]{images/s16_base_left}}
%	\qquad
%	\subfloat[width=0.45\linewidth][Right trials.]{\includegraphics[width=0.45\linewidth]{images/s16_base_Right}}
%	\\
	\subfloat[width=0.7\linewidth][Original signal.]{\includegraphics[width=0.7\linewidth]{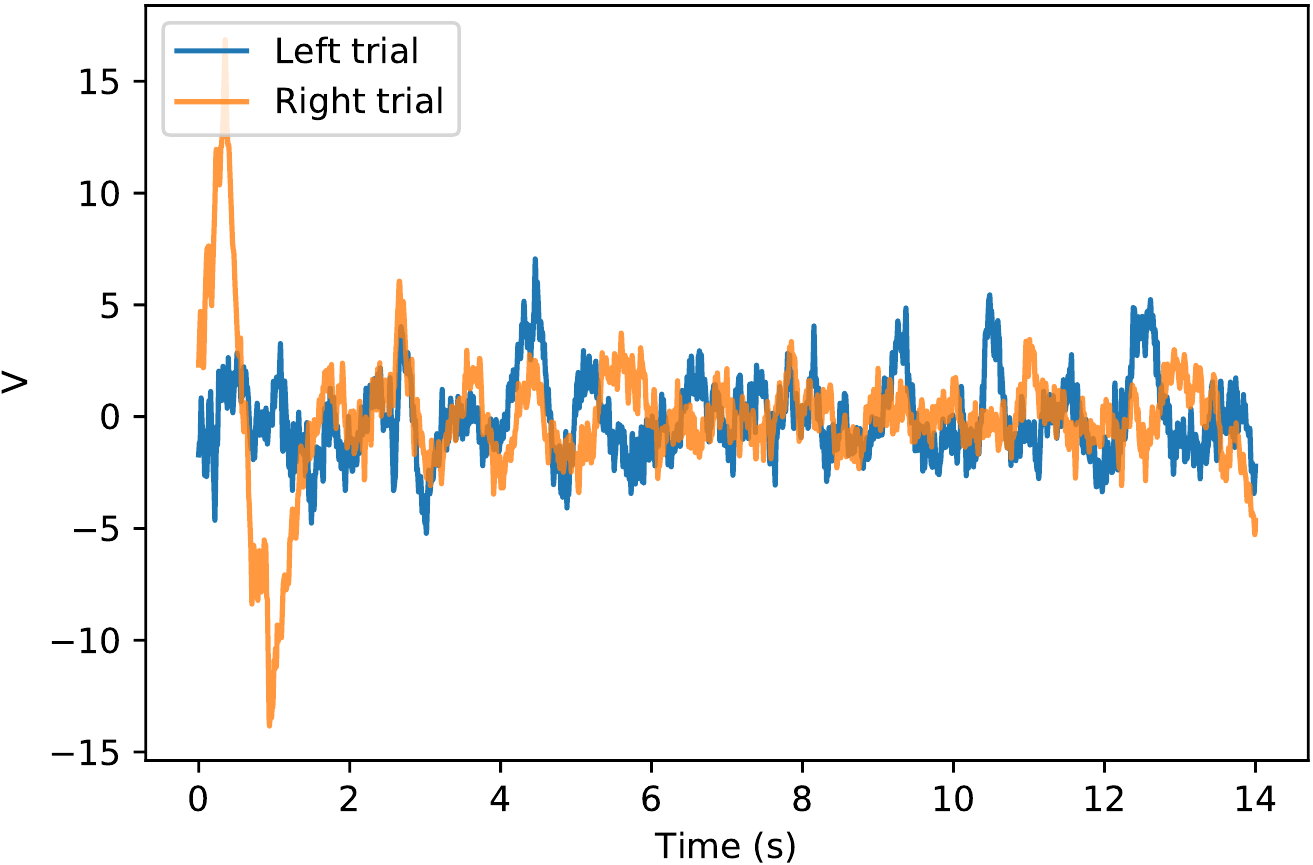}}
	\\
	\subfloat[width=0.33\linewidth][Trend.]{\includegraphics[width=0.31\linewidth]{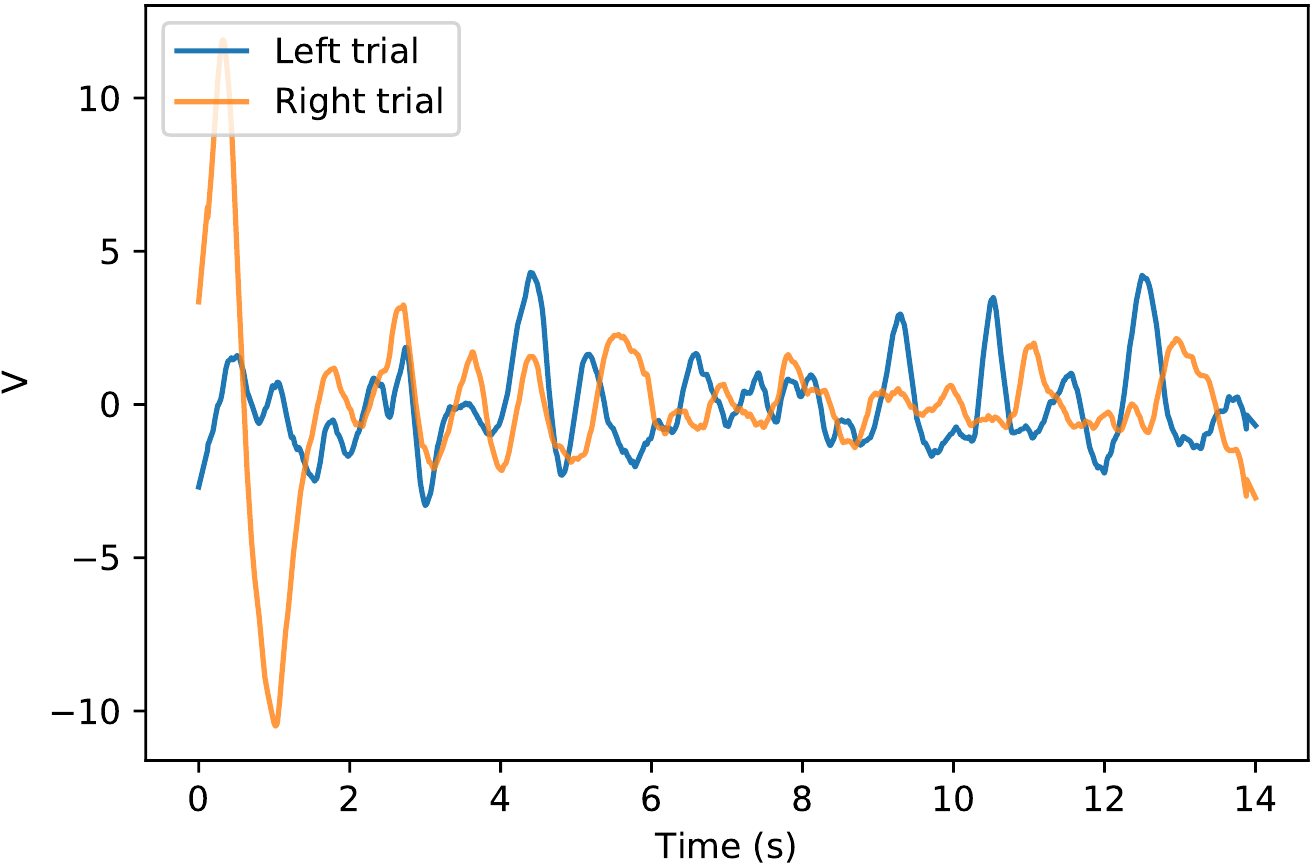}}
	\subfloat[width=0.33\linewidth][Stational comp.]{\includegraphics[width=0.31\linewidth]{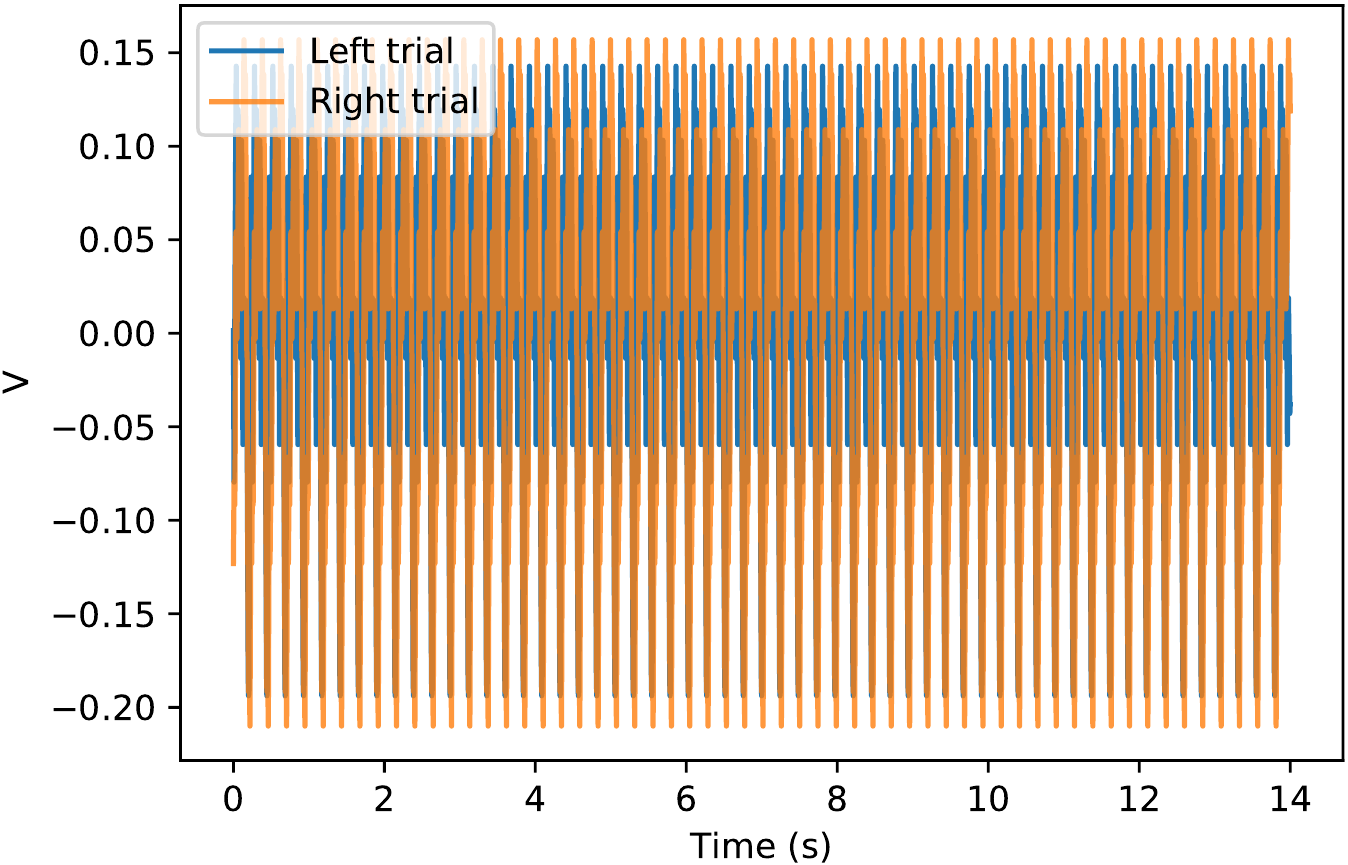}}
	\subfloat[width=0.33\linewidth][Random comp.]{\includegraphics[width=0.31\linewidth]{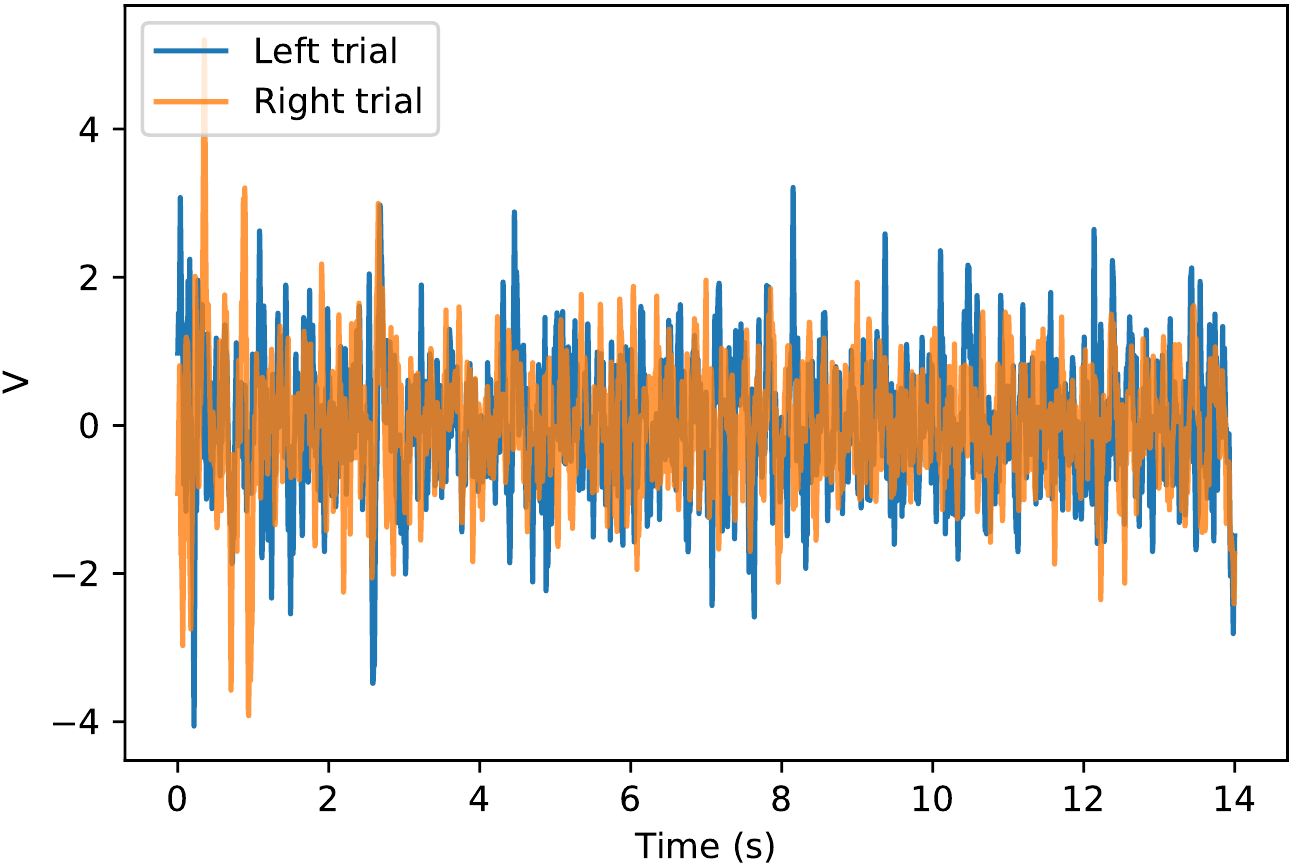}}

	\caption{Signal comparison using the average over all the waves for Subject 16. Original time series and its decomposed components.}
	\label{fig:s16_decompose}
\end{figure}

\begin{figure}
	\centering
%	\subfloat[width=0.45\linewidth][Left trials.]{\includegraphics[width=0.45\linewidth]{images/s16_base_left_diff}}
%	\qquad
%	\subfloat[width=0.45\linewidth][Right trials.]{\includegraphics[width=0.45\linewidth]{images/s16_base_Right_diff}}
%	\\
	\subfloat[width=0.7\linewidth][Differentiated original signal.]{\includegraphics[width=0.7\linewidth]{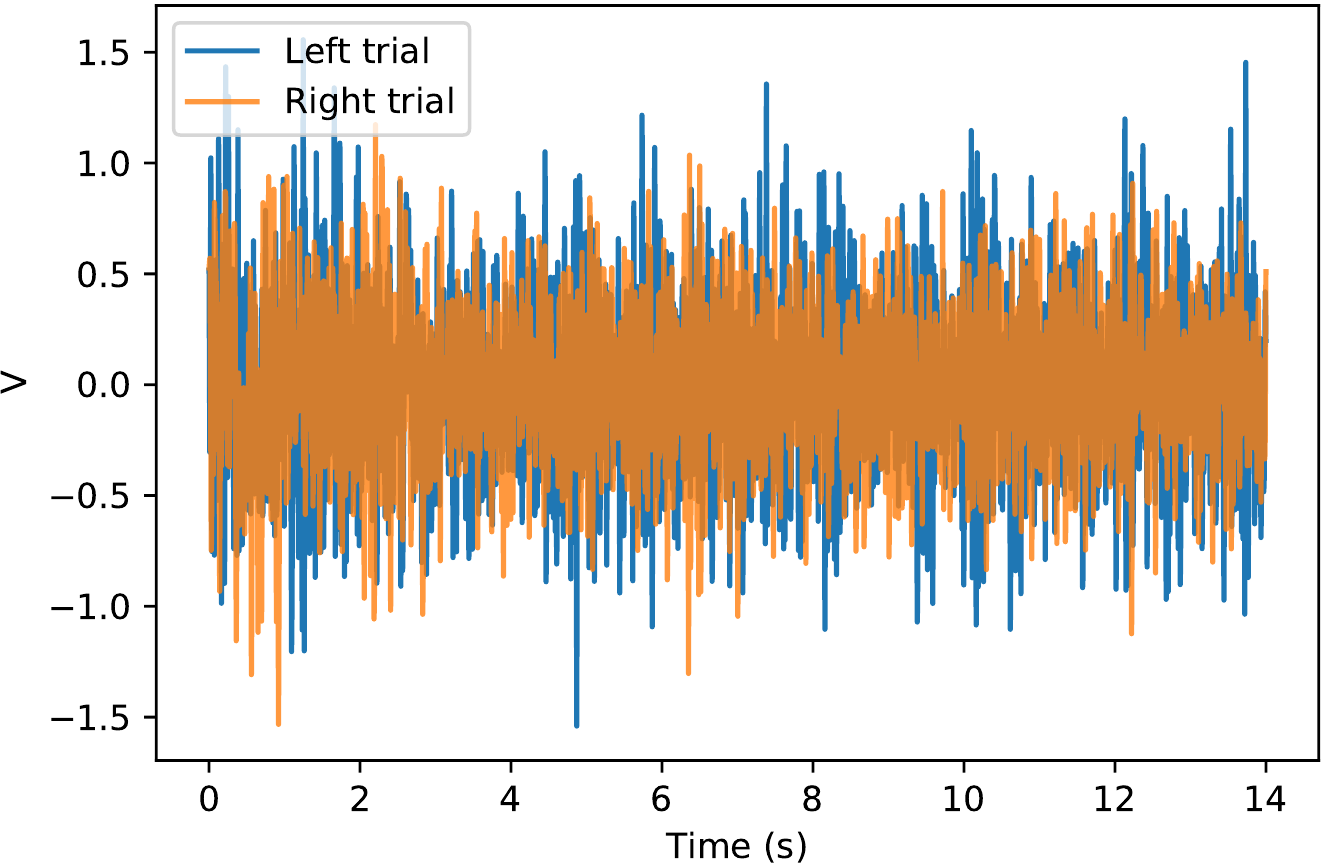}}
	\\
	\subfloat[width=0.35\linewidth][Trend.]{\includegraphics[width=0.31\linewidth]{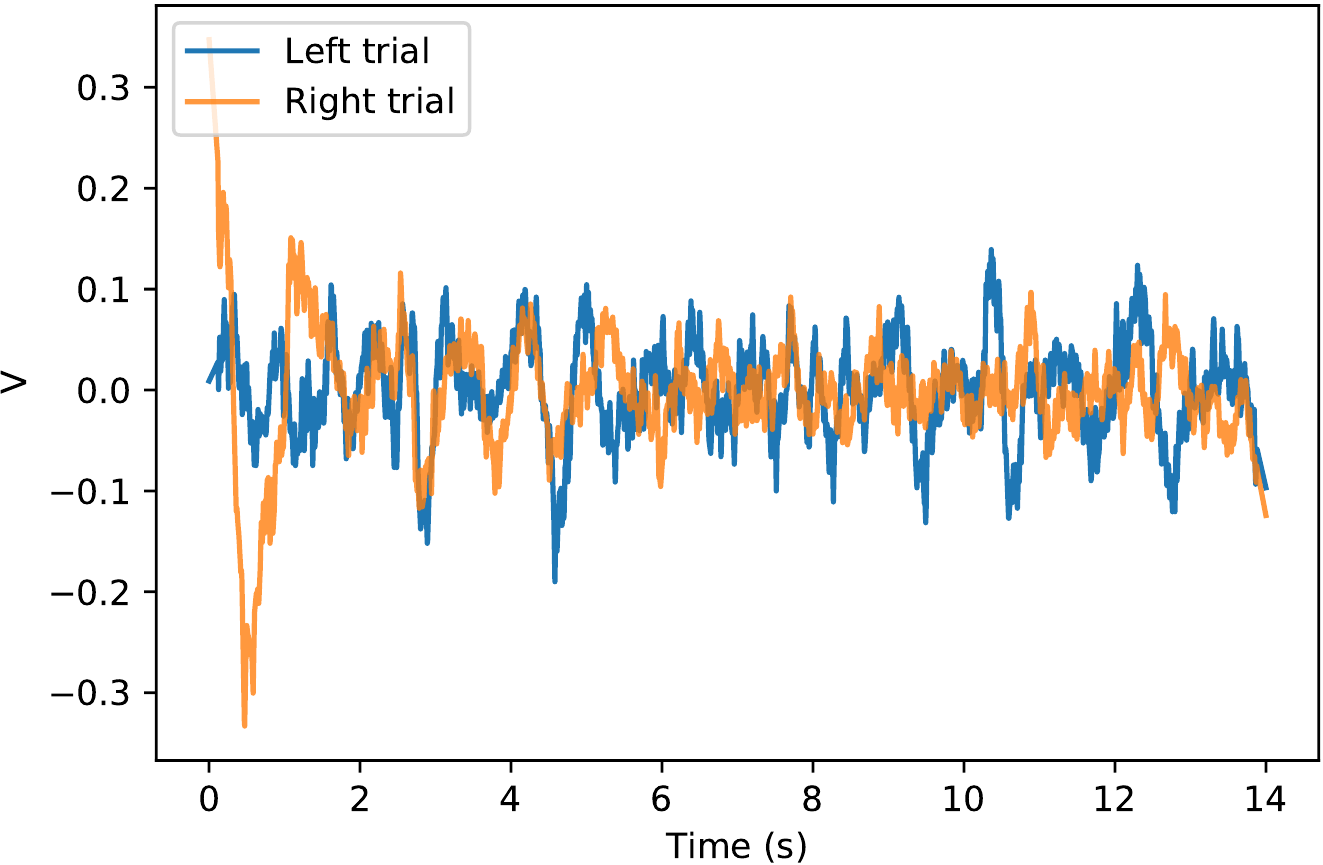}}
	\subfloat[width=0.35\linewidth][Stational comp.]{\includegraphics[width=0.31\linewidth]{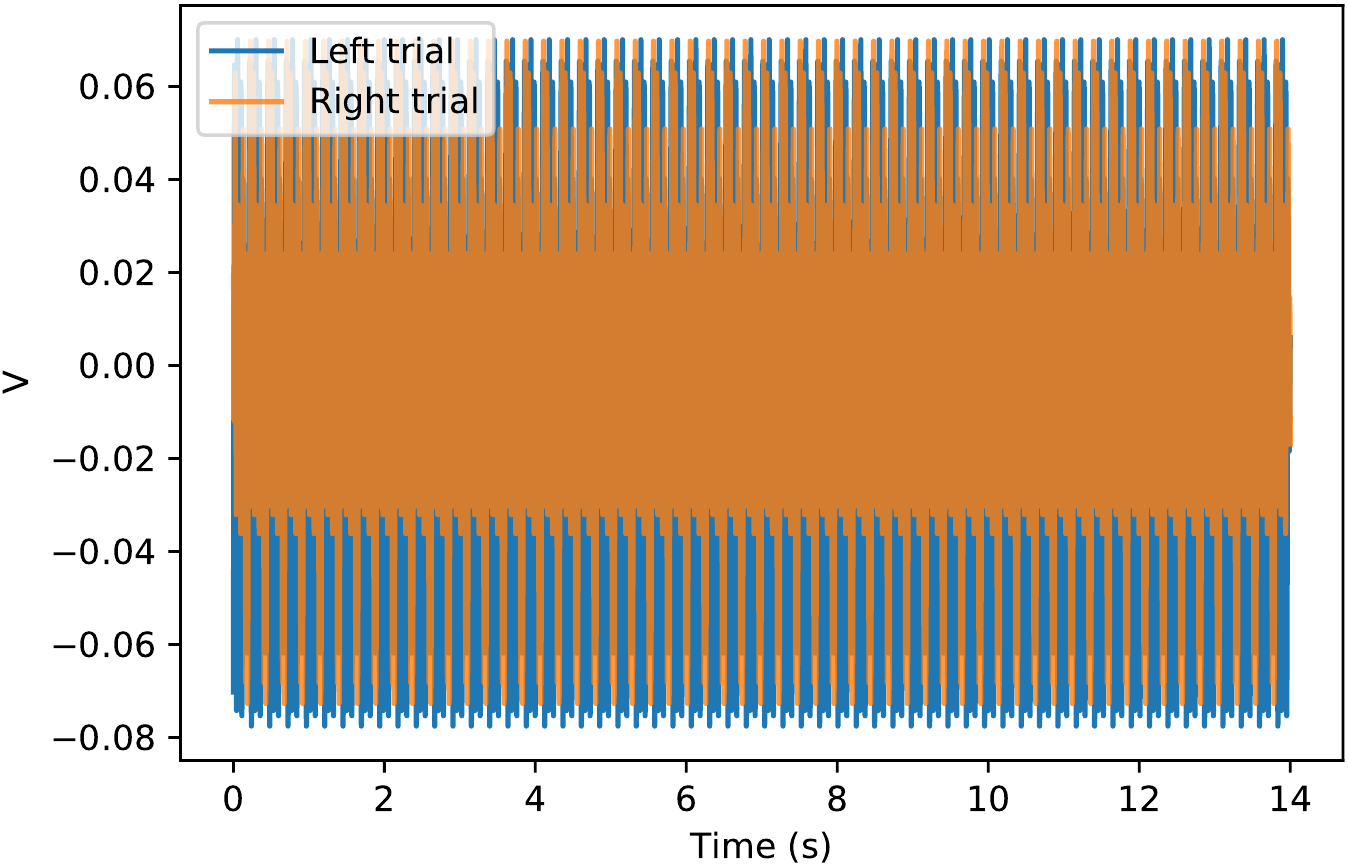}}
	\subfloat[width=0.35\linewidth][Random comp.]{\includegraphics[width=0.31\linewidth]{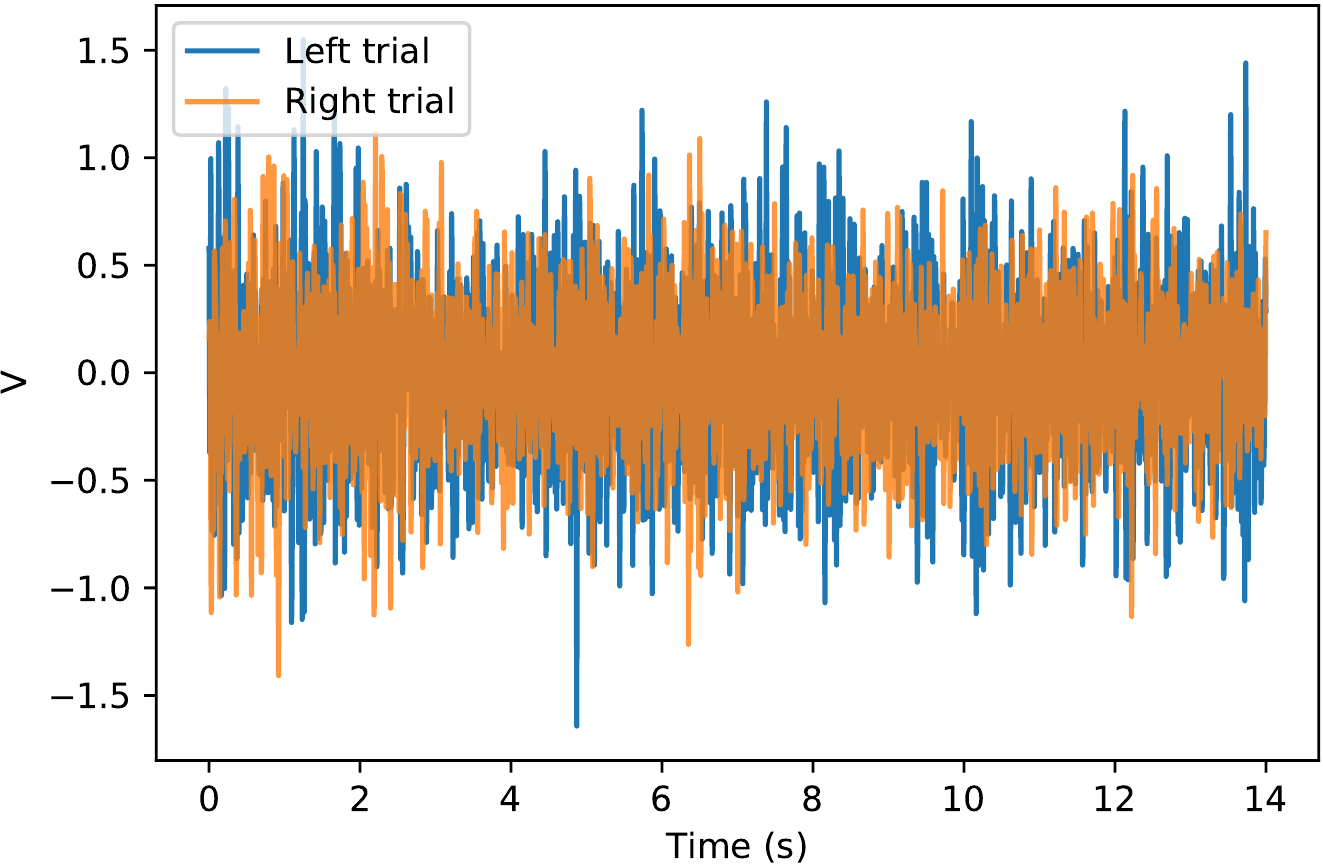}}
	
	\caption{Signal comparison using the average over all the waves for Subject 16 after being differentiated. Original differentiated time series and its decomposed components.}
	\label{fig:s16_decompose_diff}
\end{figure}

EEG data and the neuron luminescence, although different in nature, are both time series data \cite{andreas2000topics}. Temporal series are composed of three components: tendency, seasonal component and the random component \cite{hamilton1994time}, which can be observed for the average of all of our wave bands in Fig. \ref{fig:s16_decompose}. The high pass Butterworth filter used in \cite{aguilera2017signatures} is a way to remove the tendency from the time series while keeping the spikes. We have decided to do something similar to the EEG signal in order to extract the spikes in the wave bands, which are similar to the random component in the temporal series. However, instead of using the high pass Butterworth filtering, we have used a simpler procedure, the differentiation, to avoid having to tune additional parameters for the filtering. In Fig. \ref{fig:s16_decompose_diff} we show the resulting signal of the differentiation process.
% We obtained analogous results to the signal processing in the C.Elegans neural activity (Fig. \ref{fig:activityoriginal}).

\subsection{Classifiers}

In the MFF the classifiers types are LDA, QDA and KNN. They are chosen because they are ``weak" classifiers, and it is when using the aggregation process that the ``strong" decision is obtained. 

In the EMF we have decided to test two more types of classifiers: Support Vector Machines (SVM) \cite{cortes1995support} and Gaussian Processes (GP) \cite{mackay1998introduction}. These classifiers are by themselves generally more accurate than the three classifiers used in the MFF, which may increase the final accuracy of the system. However, it makes the decision making process less diverse, because the higher the accuracy for each individual classifier, the less novel info each one of them will give to take the final decision. 

\subsection{Decision Making}
In the original MFF, authors studied the effects of the arithmetic mean operator and the following aggregation functions: 
\begin{itemize}
	\item Choquet Integral and two generalizations: CF$_{min,min}$ and CF using the Hamacher T-norm.
	\item Sugeno Integral and Sugeno using the Hamacher T-norm.
\end{itemize}

The same aggregation function was used both in the frequency-phase and in the classifier phase fusion steps.

For the EMF we have considered a wider set of aggregation functions, more precisely, all the aggregation functions presented in section \ref{sec:preliminars}. That includes the classical ones, Choquet and Sugeno Integrals alongside their generalizations, OWA operators, and n-ary overlap functions. 

We have also added an extra degree of freedom to this process: the frequency fusion phase and classifier fusion phase can use a different aggregation function. We allow it because the aim in each phase is different. In the case of the frequency fusion phase, we are fusing outputs of classifiers from the same type, so their predictions are of the same nature and we are building a new collective vector. In the case of the classifier fusion phase, we are fusing different types of classifiers (even if the the outputs are normalized in the $[0,1]$ scale) and we want to make the final decision, not only building another collective vector.

\section{Experimental results for the Enhanced Multimodal Fusion Framework on the UTS MI dataset}
\label{sec:experiments}
\begin{figure}
	\centering
	\includegraphics[width=\linewidth]{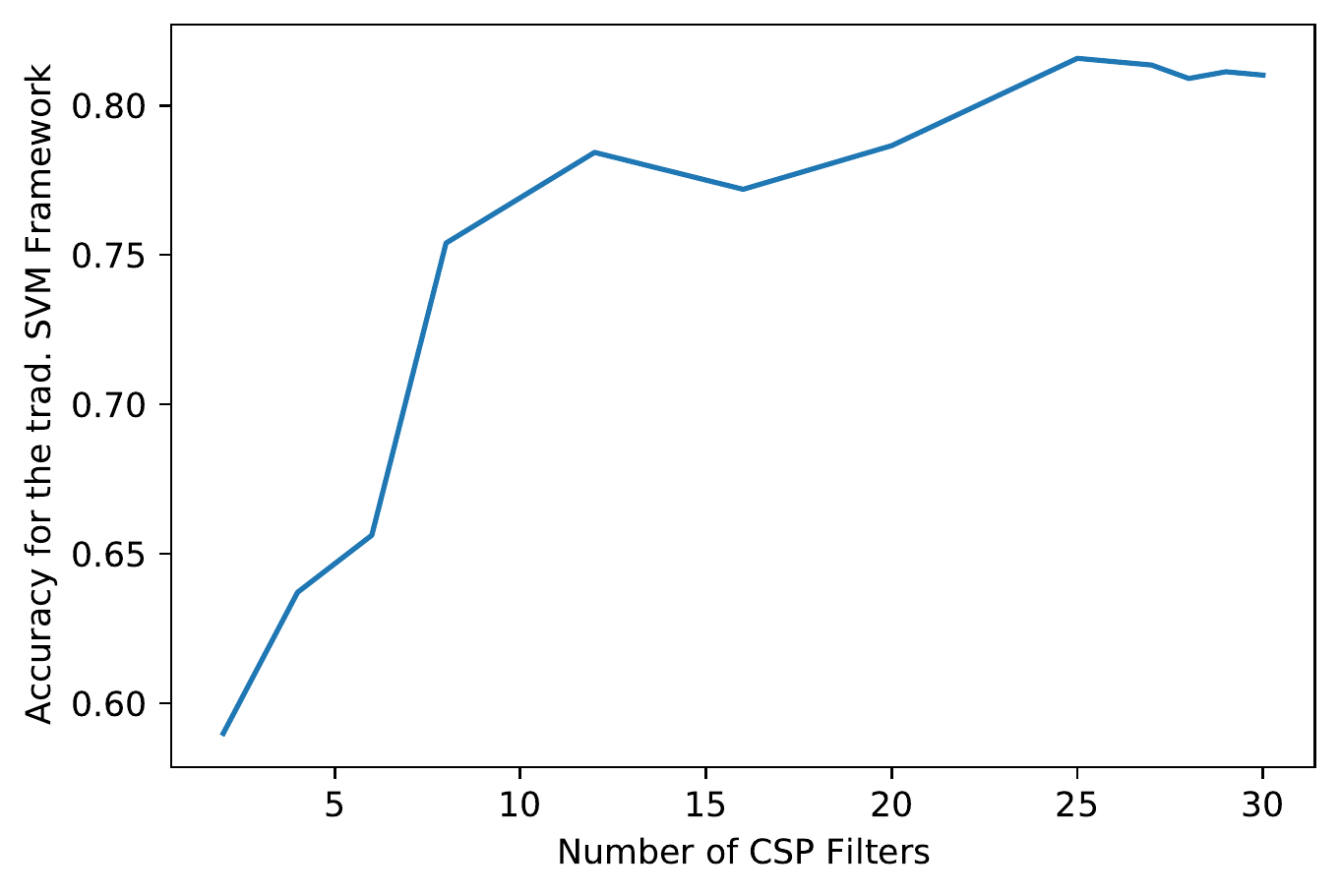}
	\caption{\note{Accuracy for the Traditional SVM BCI Framework using signal differentiation, according to a different maximum number of CSP filters for each band.}}
	\label{fig:csp_study}
\end{figure}

We have evaluated the EMF using the MI dataset collected by the \note{University of Technology Sidney} (UTS) using the same procedure as in \cite{ko2019multimodal}. This dataset consists of twenty participants. Each one of them performed a total of forty trials in which they were asked to imagine to move the left or right hand. Half of them corresponding to right, and the other half to left, consisting in a total of 800 trials. EEG data was taken from the channels C3, C4, CP3 and CP4. \note{We have used a CSP with 3, 4, 6, 15, 3 and 25 components, respectively, for the $\delta$, $\theta$, $\alpha$, $\beta$, SMR and All ($1-30Hz$) wave bands}. \note{These values have been chosen empirically (Fig. \ref{fig:csp_study}).}

%The dataset from the BCI IV competition includes 22-channel EEG signals taken from nine volunteer participants, which included not only left and right hand, but also tongue and foot. For each participant,there are 288 samples, evenly distributed among all classes.

\note{We have applied a five-fold cross validation scheme to evaluate our results: we have taken the 800 available trials, and divided it into 5 different 80/20 train-test splits, denoting the final accuracy as the mean of the accuracy for each test split. Although results here shown only show the performance of the EMF for the totality of the dataset, results for each individual subject are available online at: \url{https://github.com/Fuminides/BCI_Results}.}

\begin{table}
	\centering
	\begin{tabular}{lc}
		\toprule
		Framework 				&  Accuracy \\
		\midrule
		Trad. SVM           & $67.07$\\
		Trad. LDA          & $72.24$ \\
		Trad. QDA         & $73.82$\\
		Trad. KNN        & $68.87$\\
		Trad. GP          &  $72.47$\\
		MFF & $76.96$	\\
		EMF & $88.86$	\\
		%OEMF & $90.78$	\\
		\bottomrule
	\end{tabular}
	\caption{Performance for each BCI framework in the UTS dataset.}
	\label{tab:base_performance}
\end{table}

\begin{table}
	\centering
	\begin{tabular}{lcc}
		\toprule
		Framework 	&  Mean agg. & Best Frequency Fusion\\
		\midrule
		Diff-traditional SVM           & $79.98$ & $80.67$ \\
		Diff-traditional LDA          & $67.30$&  $74.38$\\
		Diff-traditional QDA         & $72.13$ & $83.25$\\
		Diff-traditional KNN        & $86.06$ & $86.06$\\
		Diff-traditional GP          &  $85.95$ & $86.51$\\
		\bottomrule
	\end{tabular}
	\caption{Performance for the traditional BCI frameworks using the differentiation. We compare the usage of the base aggregation (the arithmetic mean) against the best possible one.}
	\label{tab:diff_performance}
\end{table}

In Table \ref{tab:base_performance} we compared the results for the traditional framework, the MFF and our new proposal, the EMF. For the traditional framework we have used the 5 classifiers considered in this work (LDA, QDA, KNN, SVM and GP), and in the case of both the MFF, the EMF we have reported the result of the best aggregation (we will show their influence later). Looking at these results, we can observe that we have obtained a remarkable improvement using the EMF compared to any of the other frameworks, since we improve in almost $12\%$ the MFF and more than $15\%$ the best traditional framework (Trad. QDA). 

The EMF offered two main differences compared to the MFF and the traditional approaches: the differentiation phase and the enhanced frequency and classifier fusion phases. To test how much percentage of improvement comes from each one, we have computed the traditional BCI framework using the differentiation signal phase. We named this configuration ``Diff-traditional" Framework. In the traditional framework, (obviously also in the Diff-traditional), the final decision is taken fusing the information from each classifier using the arithmetic mean of all of them. Then, to see the improvement that can be achieved on the diff-traditional framework using different aggregation functions, we have tried all the aggregation functions considered in this paper and then we select the best one in terms of accuracy. The results of these tests are in Table \ref{tab:diff_performance}. In the first column we specify the classifier used for each diff-traditional framework configuration. In the second column we specify the accuracy if we use the arithmetic mean aggregation (Mean agg.), and in the third, we specify again the accuracy for the best possible aggregation function (Best Frequency Fusion). We have found that the differentiation alone is capable of boosting the performance in the case of SVM, KNN and GP more than $10\%$. In the cases were differentiation was not very successful  (LDA and QDA), the aggregation phase obtained similar winnings in terms of accuracy as the differentiation did on the other cases.

In Table \ref{tab:emf} we show the results for each pair of aggregations used in the frequency (rows) and in the classifier (columns) fusion phases using the EMF. Depending on the aggregations used, the accuracy can vary from $\sim 85\%$ to $\sim88\%$. Although there are some combinations which results in a really poor interaction between the frequency and the classifier function base. Therefore, we can conclude that in general, they provide competitive results since most of the combinations would allow to improve the results of the MFF ($76.96\%$).

Then, we also test the best possible performance of each individual wave band, which is detailed in Tab \ref{tab:single_waves}. The process is the same as in the standard EMF, but using only one wave band, so there is no classifier fusion-phase. In the first column we show the different wave bands used, in the second column the combination of classifiers that works best for this wave band, the third column is the aggregation used to fuse the information from the classifiers, and the last one shows the accuracy obtained. For example, in the case of the $\delta$ wave band, the best result is obtained using a SVM-$\delta$ and a KNN -$\delta$ classifier, and fusing their results using any aggregation function, as all of them result in the same accuracy for this case. We must stress that the $\beta$ channels alone can lead up to $\sim86\%$ accuracy using Gaussian Process and KNN classifiers and the All wave band can achieve a $89.77\%$ using only a Gaussian Process classifier (so, no aggregation process is made). \notea{The good performance of the $\beta$ band is in line with the results in \cite{krausz2003critical} and \cite{vidaurre2007study}} \note{The SMR band did not performed better than the alpha or beta wave bands, in spite of the results regarding MI reported in \cite{arroyo1993functional}}

Finally, in Table \ref{tab:itr} we show the resulting Information Transfer Rate (ITR) for the MFF, the EMF. The ITR measures the efficiency of the system, and it measured as bits/trial. It is computed using the following formulas \cite{wolpaw2002brain}:
\[
B = log_2 N + P \times log_2P+(1-P) \times log_2 \Biggl({1-P \over N-1}\Biggr) \]
\[
Q\Biggr({Trials \over Min}\Biggr) = {S \over T}
\]
\[
ITR = B * Q,
\]
where N is the number of target classes (2 in this case), S is the number of observations and P is the accuracy rate.
So, the more accuracy and the less computing time, the better this index will be. From these results we can confirm again that the EMF, using the best combination of aggregations in terms of accuracy, surpasses the MFF.
\begingroup
\setlength{\tabcolsep}{3.2pt} % Default value: 4pt
\begin{table*}[h]
	\begin{tabular}{lccccccccccccccccc}
		\toprule
		{} &  Mean & Median & Choquet & CF$_{m, m}$ & Sugeno & H. Sugeno & F-Sugeno &            Min &   Max &  C$_{F1, F2}$ &  OWA$_1$ &  OWA$_2$ &           OWA$_3$ &    CF &    GM &    SO &    HM \\
		\midrule
		Mean            & 87.08 &  85.73 &   87.08 &    86.85 &  86.85 &           86.97 &    87.42 &          87.87 & 87.87 & 87.98 & 87.08 & 87.08 & 86.52 & 87.08 & 87.98 & 88.31 & 88.31 \\
		Median          & 86.85 &  83.03 &   86.52 &    84.04 &  83.93 &           86.63 &    86.52 &          86.07 & 86.07 & 86.29 & 86.18 & 86.29 & 84.83 & 85.84 & 86.40 & 86.40 & 86.29 \\
		Choquet         & 87.08 &  85.73 &   87.75 &    86.74 &  86.74 &           87.19 &    88.09 & \textbf{88.76} & 86.97 & 88.09 & 86.63 & 87.98 & 86.63 & 87.75 & 88.31 & 88.31 & 88.31 \\
		CF$_{m, m}$     & 87.30 &  83.37 &   87.19 &    84.72 &  84.72 &           86.52 &    87.30 &          86.18 & 83.82 & 69.10 & 85.84 & 86.97 & 84.94 & 85.73 & 88.09 & 87.30 & 87.64 \\
		Sugeno          & 87.30 &  83.37 &   87.19 &    84.72 &  84.72 &           86.52 &    87.30 &          86.18 & 83.82 & 69.10 & 85.84 & 86.97 & 84.94 & 85.73 & 88.09 & 87.30 & 87.64 \\
		H. Sugeno 		& 87.64 &  85.84 &   87.30 &    86.97 &  86.97 &           86.52 &    87.98 &          87.19 & 83.82 & 87.98 & 86.52 & 87.98 & 86.85 & 87.98 & 87.98 & 88.20 & 87.87 \\
		F-Sugeno        & 87.30 &  86.18 &   87.98 &    86.97 &  86.97 &           87.30 &    88.09 & \textbf{88.76} & 86.29 & 88.09 & 86.74 & 87.98 & 87.08 & 87.75 & 88.20 & 88.09 & 87.87 \\
		Min             & 87.53 &  86.74 &   87.98 &    87.08 &  87.08 &           87.75 &    88.09 &          87.53 & 83.93 & 87.98 & 86.52 & 87.87 & 87.75 & 87.98 & 87.42 & 87.42 & 87.75 \\
		Max             & 87.53 &  86.74 &   86.97 &    84.94 &  84.94 &           85.39 &    86.63 &          83.93 & 87.53 & 86.40 & 87.75 & 85.96 & 86.85 & 86.97 & 86.97 & 86.97 & 86.63 \\
		C$_{F1, F2}$    & 87.30 &  85.84 &   87.64 &    86.52 &  86.52 &           87.19 &    13.26 &          88.54 & 64.04 & 12.13 & 86.74 & 87.75 & 86.40 & 86.18 & 88.54 & 74.16 & 88.31 \\
		OWA$_1$         & 87.08 &  85.62 &   86.40 &    85.28 &  85.28 &           86.29 &    86.40 &          85.96 & 88.31 & 86.85 & 87.30 & 86.40 & 85.96 & 86.63 & 86.85 & 86.63 & 86.63 \\
		OWA$_2$         & 87.08 &  85.73 &   87.98 &    87.08 &  87.08 &           87.08 &    88.09 &          88.31 & 85.73 & 88.31 & 86.52 & 88.20 & 86.97 & 87.53 & 88.20 & 88.20 & 87.87 \\
		OWA$_3$         & 86.52 &  83.37 &   86.85 &    84.38 &  84.38 &           85.84 &    87.75 &          88.09 & 87.30 & 87.87 & 86.18 & 87.08 & 84.72 & 85.62 & 88.09 & 88.31 & 87.64 \\
		CF              & 86.97 &  85.17 &   86.74 &    85.84 &  85.84 &           86.40 &    87.42 &          87.75 & 87.98 & 87.75 & 86.63 & 87.08 & 85.62 & 86.29 & 87.64 & 87.87 & 87.75 \\
		GM              & 87.30 &  85.96 &   87.42 &    86.74 &  86.74 &           87.08 &    87.87 &          87.64 & 87.98 & 87.87 & 86.97 & 87.64 & 86.52 & 87.30 & 87.53 & 87.64 & 87.53 \\
		SO              & 87.30 &  85.96 &   87.42 &    86.74 &  86.74 &           87.08 &    86.29 &          87.64 & 87.98 & 13.71 & 86.97 & 87.64 & 86.52 & 87.30 & 87.53 & 87.64 & 87.53 \\
		HM              & 87.30 &  86.07 &   87.53 &    86.74 &  86.74 &           87.19 &    87.75 &          88.09 & 87.87 & 87.87 & 86.74 & 87.64 & 86.52 & 87.42 & 87.64 & 87.64 & 87.87 \\
		\bottomrule
	\end{tabular}
	
	\caption{Results for the EMF in the UTS dataset.  Each row represents the aggregation function used in the frequency fusion phase, and each column the one used in the classifier fusion phase.}
	\label{tab:emf}
\end{table*}
\endgroup
\begin{table}
	\centering
	\begin{tabular}{lccc}
		\toprule
		Wave Band & Classifiers & Aggregation & Accuracy\\
		\midrule
		$\delta$ & SVM, KNN & Any &$70.44$\\
		$\theta$ & SVM, KNN, GP & CF$_{min, min}$ &$70.11$\\
		$\alpha$ & KNN, GP & Any &$81.65$\\
		$\beta$ & KNN, GP & Any &$86.17$\\
		\note{SMR} & KNN & None &$73.03$\\
		All & GP & None &$89.77$\\
		\bottomrule
	\end{tabular}
	\caption{Single wave performance, using the best classifier combination for each individual channel.}
	\label{tab:single_waves}	
\end{table}

\begin{table}
	\centering
	\begin{tabular}{lcc}
		\toprule
		Framework & Aggregation(s) & ITR (bit/min)\\
		\midrule
		MFF & Choquet Integral & 901.40\\
		EMF & Choquet, OWA$_3$ &  1710.74 \\
		\bottomrule
	\end{tabular}
	\caption{ITR comparison table.}
	\label{tab:itr}	
\end{table}

\begingroup
\setlength{\tabcolsep}{4pt} % Default value: 4pt
\begin{table}
	\centering
	\begin{tabular}{llllc}
		\toprule
		Wave sets &        Classifiers &   Freq. Agg. &  Class. Agg. &  Accuracy \\
		\midrule
		$\alpha$, $\beta$, All &  QDA, KNN, GP &             H. Sugeno &               GM &  90.78 \\
		$\beta$, All &       GP &  Any* &             Any* &  90.44 \\
		$\theta$, $\alpha$, $\beta$ &           LDA, QDA, KNN, GP &             H. Sugeno &             Mean &  90.44 \\
		$\theta$, $\alpha$, $\beta$, All &       QDA, KNN, GP &         F-Sugeno &         GM &  90.44 \\
		$\alpha$, $\beta$, All &  QDA, KNN, GP &          OWA1 &         GM &  90.33 \\
		$\alpha$, $\beta$, All &            QDA, KNN, GP &             Mean &              SO &  90.22 \\
		$\theta$, $\alpha$, $\beta$ &  LDA, QDA, KNN, GP &              C$_{F1, F2}$ &           Mean &  90.22 \\
		
		$\beta$, All &  QDA, KNN &             C$_{F1, F2}$ &              Min &  90.11 \\
		$\theta$, $\beta$, All &            KNN, GP &           CF &             Choquet &  90.11 \\
		$\delta$, $\alpha$, $\beta$, All &            KNN, GP &           Min &             Mean &  90.11 \\
		\bottomrule
	\end{tabular}
	\caption{Top 10 configurations in terms of accuracy for the UTS dataset. Including aggregation functions used and channels used. * Any aggregation gave the same result.}
	\label{tab:results_classifiers}
\end{table}
\endgroup

\subsection{Optimal Enhanced Motor Fusion: Classifier and wave band selection in the Enhanced Motor Fusion Framework}

Diversity is an important part of an ensemble of classifiers \cite{kuncheva2003measures}. As it seems logical, if a set of classifiers always give the same output, its combination will not give us new information. We are using a total of thirty classifiers in the EMF: five for each band (LDA, QDA, KNN, SVM and GP) and a total of six bands ($\delta$, $\theta$, $\alpha$, $\beta$, SMR, All). To measure the diversity of this system we have used the Q-statistic \cite{kuncheva2000independence}, whose result is 0.99 meaning that the diversity in the output of these classifiers is scarce, which might be impacting the performance of the system. 

One simple way to improve the system diversity, and reduce the Q-statistic value, is to erase components of the system, which is likely to improve the accuracy. Since there are only 30 classifiers, we can compute all the possible configurations of the system to see which combinations of classifiers and which wave bands are the most suited \note{according to their accuracy}. After computing all the possible 1860 combinations of wave bands and classifiers, we have determined which are the top configurations in terms of accuracy, which are shown in Table \ref{tab:results_classifiers}. As suspected, reducing the number of components has a good impact on the performance, since we can see that the best configuration is able to improve the EMF in almost $2\%$. It also improves the ITR compared to the previous tests, obtaining a value of $2007.47$bit/min. We have named this optimal configuration of the EMF ``Optimal Enhanced Motor Fusion" (OEMF).

\section{Comparison on the BCI competition IV 2a dataset}
\label{sec:bciiv_dataset}

In this section we discuss the behaviour of our new approaches in the BCI competition IV dataset 2a, which is detailed in \cite{brunner2007spatial}. This dataset consists of four classes of imaginary tasks: tongue, foot, left-hand and right-hand performed by 9 volunteers. For each task, 22 EEG channels were collected. There is a total of 288 trials for each participants, equally distributed among the 4 classes.

For our experimental setup, we have used 4 out of the 22 channels available (8, 12, 14, 18). As features, we have used the FFt to obtain the $\delta$, $\theta$, $\alpha$, $\beta$, SMR and All, and \note{the CSP filters are the same in Section \ref{sec:experiments}}. \note{From each subject, we have generated twenty partitions of the 288 trials consisting of 50\% train (144 trials) and 50\% test (144 trials) chosen randomly. Since we have 9 subjects, this produces a total of 180 different datasets. We do this in order to compute a population of accuracies for each classifier, which allow us to calculate the statistical significances among them.}

We have studied this dataset from two different perspectives:
\begin{enumerate}
	\item Taking the four classes to perform the classification task.
	\item Taking only the left and right hand classes, so that the classification task is the same as in the previous section.
\end{enumerate}
\note{We have not reported the accuracy for each individual class, as the differences among them were not significant.}

We show the results for the four-classes case in Fig. \ref{fig:accuracy_comparison_2a}.  We compare the same set of traditional frameworks as in the UTS dataset, alongside the MFF, the EMF and the OEMF. We can observe that the performance of EMF is $83.15\%$, using the C$_{F1,F2}$ and the harmonic mean functions in the frequency and classifier based fusion phases, respectively, which is better than that of both the traditional framework with all the classifiers and the MFF. The performances of all the pairs of combinations of the aggregation functions for the EMF can be seen in Table \ref{tab:emf_bci_iv}. When applying the OEMF, we obtain a $85.62\%$ of accuracy using the LDA, QDA and KNN classifiers, and all the wave bands available. This implies a notable enhancement on the results of the EMF, which confirms the suitability of this procedure to tackle this task. We have also tested the statistical the results for all the BCI frameworks using the multi-population Friedman test, comparing the population of 180 accuracies for each framework. The result for the multi population comparison gave us a \textit{P}-value $<0.001$. Then, we applied the Nemenyi post-hoc to confirm that both the EMF and the OEMF are statistically better than the rest of the methods, although they are not statistically different one from another. 

%We obtained significant statistical differences for the multi-population comparison using the Friedman test, and the \textit{P}-values for the Nemenyi post-hoc test are also in Fig. \ref{fig:accuracy_comparison_2a}.

In Fig. \ref{fig:accuracy_comparison_2a_binary} we show the results for the two classes problem, using the same frameworks as in the four classes case. As expected, the accuracy of all classifiers improves. The EMF increases almost $3\%$ with respect to the four classes problem. In this scenario, the best pair of aggregations for the EMF is the Hamacher Sugeno and the Sin overlap. The rest of the results for each pair of aggregations can be seen in Tab. \ref{tab:emf_bci_iv_left_right}.

The best configuration obtained by the OEMF (is using all the possible wave bands and QDA and KNN classifiers). This configuration presents a huge improvement of more than $12\%$ over the EMF, reaching a total of $97.60\%$. We have also studied the statistical differences among the frameworks, using an analogue procedure to the one performed in the four-classes problem. We have found that the EMF and the OEMF are statistically better than the rest of the frameworks. In this case, the OEMF is also statistically better than the EMF.

\note{We have also studied} the impact of the pair of aggregations to be used in our new framework, since it seems to have an impact on the obtained results. There are some combinations which tend to perform always near the best possible performance, and others that do the opposite. Our results show that usually, the best combination is using a Sugeno/Choquet integral in the frequency fusion phase and an overlap function in the classifier fusion phase, which can be appreciated in Tables \ref{tab:emf}, \ref{tab:emf_bci_iv}, \ref{tab:emf_bci_iv_left_right}.

\note{Finally, we have compared our results in the four classes task with three other kinds of BCI frameworks in Table \ref{tab:others}. In \cite{mishra2018novel} the authors combine Riemannian geometry with a Linear SVM. In \cite{8452948} the authors optimize the time interval for each subject to obtain the optimal set of features to discriminate between tasks, and then apply a bio-inspired algorithm to optimize the CSP features and SVM parameters. And the authors in \cite{RAZI201914} use the Dempster–Shafer theory to fuse the output of different LDA classifiers. We found that the results obtained by the EMF are higher than the other three BCI frameworks.}

\begin{figure}
	\centering
	%\includegraphics[width=\linewidth]{images/2a_full_plot}
	%\\ \phantom{Ghosteen is the best album in the 21st century and Joan of Arc is the best catholic saint.} 
	\begin{tabular}{lc}
		\toprule
		Framework 				&  Accuracy \\
		\midrule
		Trad. SVM           & $60.67 \pm 4.23  $\\
		Trad. LDA         & $69.94 \pm 4.05$\\
		Trad. QDA          & $47.93 \pm 2.82	$\\
		Trad. KNN        & $60.56	\pm 4.20$\\
		Trad. GP          &  $53.04 \pm 4.60$\\
		MFF & $67.96 \pm 2.19	$\\
		EMF & $83.15 \pm 3.02$\\
		OEMF & $85.40 \pm  3.03$	\\
		\bottomrule
	\end{tabular}
	\\ \phantom{Ghosteen is the best album in the 21st century and Joan of Arc is the best catholic saint.}
	\begingroup
	\setlength{\tabcolsep}{3pt} % Default value: 4pt
	\begin{tabular}{lllllllll}
		\toprule
		\textit{P}-values & SVM 	&  LDA  &  QDA  & KNN  &  GP  &  MFF&  EMF &  OEMF \\
		\midrule	
		SVM  			  & $-$		& {}  	& {} 	& {}  & {}  & {}    &	{}  &   {}	 \\
		LDA   			  & $*$		& $-$ 	& {} 	& {}  & {}  & {}    &	{}  &   {}	 \\
		QDA  			  & $+$		& $+$	& $-$	& {}  & {}  & {}    &	{}  &   {}	 \\
		KNN  			  & $.9$		& $+$	& $*$	& $-$ & {}  & {} 	&	{}  &   {}	 \\
		GP  			  & $+$		& $+$	& $.28$	& $+$ & $-$ &  {}	&   {}  &   {}	 \\
		MFF 			  & $*$		& .9	& $*$	& $*$ & $*$ &  $-$ 	&   {}	&   {}   \\
		EMF  			  & $*$		& $*$	& $*$	& $*$ & $*$	& $*$   &   $-$ &	{}   \\
		OEMF  			  & $*$		& $*$	& $*$	& $*$ & $*$ &  $*$  & 	$.84$ &	$-$	 \\
		\bottomrule
	\end{tabular}

	\endgroup
	\caption{Accuracy comparison for the BCI Competition IV dataset 2a using the four classes. $*$ represents a p-value less than 0.001, favouring the framework in the row. $+$ represents a p-value less than 0.001, favouring the framework in the column.}
	\label{fig:accuracy_comparison_2a}
\end{figure}

\begin{figure}
	\centering
	%\includegraphics[width=\linewidth]{images/2a_binary_full_plot}
%	\\ \phantom{Ghosteen is the best album in the 21st century and Joan of Arc is the best catholic saint.} 
	\begin{tabular}{lc}
		\toprule
		Framework 				&  Accuracy \\
		\midrule
		Trad. SVM           & $73.97 \pm 4.79$  \\
		Trad. LDA          & $80.75 \pm 4.42$	\\
		Trad. QDA         & $67.15 \pm 4.49$\\
		Trad. KNN        & $71.41 \pm 4.90$	\\
		Trad. GP          &  $66.84 \pm 4.32$\\
		MFF & $81.15 \pm 1.32$	\\
		EMF & $85.83 \pm 1.68$	\\
		OEMF & $97.60 \pm 1.03$	\\
		\bottomrule
	\end{tabular}
\\ \phantom{Ghosteen is the best album in the 21st century and Joan of Arc is the best catholic saint.} 
	\begingroup
	\setlength{\tabcolsep}{3pt} % Default value: 4pt
	\begin{tabular}{lllllllll}
		\toprule
		\textit{P}-values & SVM 	&  LDA  &  QDA  & KNN  &  GP  &  MFF&  EMF &  OEMF \\
		\midrule	
		SVM  			  & $-$		& {}  	& {} 	& {}  & {}  & {}    &	{}  &   {}	 \\
		LDA   			  & $*$		& $-$ 	& {} 	& {}  & {}  & {}    &	{}  &   {}	 \\
		QDA  			  & $+$		& $+$	& $-$	& {}  & {}  & {}    &	{}  &   {}	 \\
		KNN  			  & $+$		& $+$	& $.04$	& $-$ & {}  & {} 	&	{}  &   {}	 \\
		GP  			  & $+$		& $+$	& $*$	& $+$ & $-$ &  {}	&   {}  &   {}	 \\
		MFF 			  & $*$		& .9	& $*$	& $*$ & $*$ &  $-$ 	&   {}	&   {}   \\
		EMF  			  & $*$		& $*$	& $*$	& $*$ & $*$	& $*$   &   $-$ &	{}   \\
		OEMF  			  & $*$		& $*$	& $*$	& $*$ & $*$ &  $*$  & 	$.01$ &	$-$	 \\
		\bottomrule
	\end{tabular}
	\endgroup
	
	\caption{Accuracy comparison for the BCI Competition IV dataset 2a using only the left-hand and right-hand classes. $*$ represents a p-value less than 0.001, favouring the framework in the row. $+$ represents a p-value less than 0.001, favouring the framework in the column.}
	\label{fig:accuracy_comparison_2a_binary}
\end{figure}

\begingroup
\setlength{\tabcolsep}{3.2pt} % Default value: 4pt
\begin{table*}
	\begin{tabular}{lccccccccccccccccc}
		\toprule
		{} &  Mean & Median & Choquet & CF$_{m,m}$ & Sugeno & H. Sugeno & F-Sugeno &   Min &   Max &  C$_{F1,F2}$ &  OWA$_1$ &  OWA$_2$ &  OWA$_3$ &    CF &    GM &    SO &             HM \\
		\midrule
		Mean            & 78.38 &  70.69 &   77.48 &    69.10 &  69.10 &           70.70 &    77.46 & 79.36 & 78.52 & 79.40 & 77.13 & 76.21 & 73.74 & 75.59 & 81.55 & 82.49 &          81.20 \\
		Median          & 78.04 &  70.12 &   76.71 &    67.61 &  67.60 &           69.41 &    76.56 & 76.88 & 76.57 & 77.61 & 76.49 & 75.15 & 73.58 & 75.39 & 78.58 & 78.58 &          77.80 \\
		Choquet         & 77.46 &  70.09 &   77.98 &    68.43 &  68.43 &           70.69 &    78.60 & 81.41 & 76.44 & 79.92 & 75.90 & 77.46 & 73.76 & 74.93 & 82.57 & 82.79 &          82.40 \\
		CF$_{m,m}$      & 78.01 &  70.12 &   76.71 &    67.61 &  67.60 &           69.41 &    77.46 & 73.50 & 72.93 & 64.30 & 75.30 & 74.96 & 73.58 & 75.28 & 78.58 & 80.34 &          77.80 \\
		Sugeno          & 78.01 &  70.12 &   76.71 &    67.61 &  67.60 &           69.41 &    77.46 & 73.52 & 72.93 & 64.30 & 75.30 & 74.96 & 73.58 & 75.28 & 78.58 & 80.34 &          77.80 \\
		H. Sugeno 		& 76.08 &  68.73 &   77.47 &    65.92 &  65.85 &           70.23 &    78.00 & 77.03 & 70.55 & 79.58 & 71.93 & 76.26 & 73.12 & 73.25 & 80.22 & 82.50 &          79.33 \\
		F-Sugeno        & 76.88 &  69.70 &   77.93 &    68.01 &  68.01 &           70.48 &    78.80 & 82.09 & 75.24 & 80.10 & 74.92 & 77.69 & 73.55 & 74.47 & 82.84 & 82.71 &          82.72 \\
		Min             & 74.08 &  67.19 &   76.70 &    65.13 &  65.06 &           68.80 &    78.02 & 81.58 & 66.08 & 79.56 & 69.92 & 77.21 & 71.58 & 71.60 & 81.28 & 81.28 &          81.54 \\
		Max             & 77.74 &  67.61 &   74.15 &    65.70 &  65.70 &           67.01 &    71.96 & 64.58 & 78.12 & 73.81 & 75.68 & 69.27 & 70.32 & 73.67 & 76.02 & 76.01 &          73.16 \\
		C$_{F1,F2}$     & 77.43 &  70.16 &   77.93 &    68.55 &  68.55 &           70.63 &    78.08 & 82.71 & 77.11 &  2.69 & 75.95 & 77.42 & 73.74 & 74.96 & 82.64 & 71.93 & \textbf{83.15} \\
		OWA$_1$         & 79.11 &  70.65 &   76.33 &    68.66 &  68.66 &           69.97 &    75.55 & 73.78 & 79.22 & 76.87 & 77.65 & 73.79 & 73.23 & 75.90 & 78.18 & 78.28 &          76.85 \\
		OWA$_2$         & 76.83 &  69.53 &   77.87 &    67.70 &  67.70 &           70.48 &    78.80 & 82.00 & 74.33 & 80.09 & 74.40 & 77.77 & 73.58 & 74.52 & 82.52 & 82.63 &          82.34 \\
		OWA$_3$         & 78.29 &  70.28 &   77.26 &    68.36 &  68.36 &           70.13 &    77.40 & 79.18 & 77.23 & 78.60 & 76.81 & 75.84 & 73.91 & 75.67 & 79.87 & 79.75 &          79.91 \\
		CF              & 78.78 &  70.52 &   76.54 &    68.82 &  68.82 &           69.75 &    77.18 & 76.80 & 77.87 & 79.53 & 77.35 & 74.20 & 73.36 & 75.98 & 80.41 & 82.26 &          79.93 \\
		GM              & 78.05 &  70.51 &   77.72 &    69.05 &  69.05 &           70.69 &    78.25 & 81.52 & 77.81 & 79.60 & 76.73 & 77.10 & 73.78 & 75.47 & 80.95 & 81.06 &          81.53 \\
		SO              & 78.05 &  70.51 &   77.72 &    69.05 &  69.05 &           70.69 &    78.25 & 81.52 & 77.81 &  3.34 & 76.73 & 77.10 & 73.78 & 75.47 & 80.95 & 81.06 &          81.53 \\
		HM              & 77.73 &  70.41 &   77.85 &    68.88 &  68.88 &           70.62 &    78.47 & 81.62 & 76.55 & 79.68 & 76.33 & 77.46 & 73.76 & 75.29 & 81.31 & 81.27 &          81.66 \\
		\bottomrule
	\end{tabular}
	
	\caption{Performance in the EMF for each pair of aggregations, for the BCI competition IV 2a dataset using the four classes. Each row represents the aggregation function used in the frequency fusion phase, and each column the one used in the classifier fusion phase.}
	\label{tab:emf_bci_iv}

\end{table*}

\begin{table*}
	\begin{tabular}{lccccccccccccccccc}
		\toprule
		{} &  Mean & Median & Choquet & CF$_{m,m}$  & Sugeno & H. Sugeno & F-Sugeno &   Min &   Max &  C$_{F1,F2}$ &  OWA1 &  OWA2 &  OWA3 &    CF &    GM &             SO & HM \\
		\midrule
		Mean            & 81.24 &  77.81 &   81.24 &    77.87 &  77.87 &           79.13 &    81.27 & 82.28 & 82.28 & 83.17 & 80.31 & 81.24 & 79.37 & 80.41 & 83.54 &          84.83 & 83.46 \\
		Median          & 78.89 &  75.43 &   79.42 &    76.31 &  76.31 &           79.00 &    79.51 & 79.92 & 79.92 & 78.87 & 78.52 & 79.20 & 77.19 & 78.36 & 79.15 &          79.15 & 79.39 \\
		Choquet         & 81.24 &  77.81 &   82.17 &    78.37 &  78.37 &           80.14 &    82.57 & 83.67 & 79.48 & 83.92 & 78.74 & 82.53 & 79.78 & 80.56 & 84.98 &          85.21 & 84.88 \\
		CF$_{m,m}$      & 80.37 &  77.18 &   81.54 &    77.01 &  76.99 &           79.14 &    83.48 & 79.95 & 75.96 & 56.37 & 76.59 & 80.85 & 79.12 & 79.54 & 81.67 &          85.67 & 81.64 \\
		Sugeno          & 80.37 &  77.18 &   81.54 &    77.01 &  76.99 &           79.14 &    83.48 & 79.94 & 75.96 & 56.37 & 76.59 & 80.85 & 79.12 & 79.54 & 81.67 &          85.67 & 81.62 \\
		H. Sugeno 		& 81.39 &  78.11 &   82.15 &    78.11 &  78.11 &           80.28 &    83.26 & 81.34 & 73.98 & 84.83 & 76.24 & 82.04 & 80.37 & 81.35 & 83.33 & \textbf{85.83} & 82.67 \\
		F-Sugeno        & 81.45 &  78.05 &   82.63 &    78.58 &  78.58 &           80.41 &    83.11 & 84.00 & 78.49 & 84.33 & 78.17 & 82.88 & 80.12 & 80.81 & 85.24 &          85.39 & 85.19 \\
		Min             & 81.69 &  78.06 &   83.58 &    77.99 &  77.99 &           80.37 &    83.98 & 85.40 & 73.98 & 84.83 & 76.22 & 83.56 & 80.68 & 81.48 & 85.83 &          85.83 & 85.73 \\
		Max             & 81.69 &  78.06 &   78.13 &    75.49 &  75.50 &           74.77 &    76.78 & 73.98 & 85.40 & 79.74 & 82.97 & 76.08 & 78.78 & 80.98 & 80.17 &          80.17 & 78.69 \\
		C$_{F1,F2}$     & 80.42 &  77.16 &   82.29 &    77.89 &  77.89 &           80.12 &    18.90 & 85.05 & 68.37 & 16.80 & 80.27 & 82.75 & 79.01 & 79.33 & 85.04 &          44.28 & 85.49 \\
		OWA$_1$         & 80.79 &  77.70 &   78.79 &    75.88 &  75.88 &           77.02 &    78.24 & 78.13 & 83.17 & 79.38 & 81.73 & 77.70 & 78.25 & 80.09 & 79.74 &          79.73 & 78.92 \\
		OWA$_2$         & 81.24 &  77.81 &   82.58 &    78.47 &  78.47 &           80.49 &    82.96 & 83.74 & 77.78 & 84.20 & 77.49 & 82.92 & 80.06 & 80.55 & 85.08 &          85.33 & 85.03 \\
		OWA$_3$         & 79.93 &  76.58 &   80.37 &    77.33 &  77.33 &           79.36 &    80.73 & 81.32 & 80.29 & 80.94 & 78.98 & 80.49 & 78.36 & 79.47 & 80.64 &          80.84 & 81.04 \\
		CF              & 80.89 &  77.74 &   79.67 &    76.83 &  76.83 &           77.72 &    81.67 & 81.37 & 81.63 & 83.79 & 80.55 & 79.70 & 78.64 & 80.27 & 83.52 &          84.75 & 83.54 \\
		GM              & 81.28 &  77.89 &   81.63 &    77.95 &  77.95 &           79.44 &    82.16 & 85.46 & 81.44 & 83.49 & 80.01 & 82.09 & 79.45 & 80.37 & 84.80 &          85.02 & 85.42 \\
		SO              & 81.28 &  77.89 &   81.63 &    77.95 &  77.95 &           79.44 &    78.49 & 85.46 & 81.44 & 16.51 & 80.01 & 82.09 & 79.45 & 80.37 & 84.80 &          85.02 & 85.42 \\
		HM              & 81.35 &  78.00 &   82.23 &    78.02 &  78.02 &           79.78 &    82.81 & 85.75 & 80.56 & 83.70 & 79.77 & 82.80 & 79.63 & 80.52 & 85.07 &          85.03 & 85.45 \\
		
		\bottomrule
	\end{tabular}

	\caption{Performance in the EMF for each pair of aggregations, for the BCI competition IV 2a dataset using only the left-right classes. Each row represents the aggregation function used in the frequency fusion phase, and each column the one used in the classifier fusion phase.}
	\label{tab:emf_bci_iv_left_right}
	
\end{table*}

\endgroup
\begin{table}
	\centering
	\begin{tabular}{cc}
		\toprule
		Framework &  Accuracy\\ 
		\midrule
		EMF &  $83.15\%$ \\ 
		%MFF \cite{ko2019multimodal} &  $81.15\%$ \\ 
		KLRRM + LSVM  \cite{mishra2018novel} &  $74,43\%$ \\
		CSP/AM-BA-SVM \cite{8452948} &  $78.55\%$ \\ 
		Dempster-Shafter  \cite{RAZI201914} &  $81.93\%$ \\ 
		\bottomrule 
	\end{tabular} 
	\caption{\note{Comparison of different BCI frameworks in the four classes task.}}
	\label{tab:others}
\end{table}
\section{Conclusions}
\label{sec:conclusions}
In this paper, we proposed a new BCI framework, named Enhanced Motor Fusion framework (EMF). The EMF proposes three new ideas to improve the Multimodal Fuzzy Fusion (MFF) BCI system performance: 
\begin{enumerate}
	\item A signal differentiation step.
	\item Both a new wave band (SMR) and new types of classifiers: SVM and GP.
	\item The usage of different aggregation functions in each phase in the decision making process.
\end{enumerate}

Furthermore, we have also enlarged the number of aggregation functions used in the decision making process, like C$_{F1,F2}$, overlap functions and the OWA operators. We observe that the best results are obtained using a Sugeno/Choquet integral in the first phase and an overlap function in the second phase. 
%Finally, we have studied the effects of each set of features in the system, choosing the different wave bands and the different types of classifiers in order to obtain the optimal combination of elements in the system. 
Moreover, we have also proposed an optimized version of the EMF by obtaining an optimal combination of the different wave bands and the different types
of classifiers.
The performance of our new approaches is tested on two BCI datasets, where the suitability of our new methods is proven. Specifically, the EMF improves the results of the traditional framework as well as those of the MFF. The OEMF improves even more the obtained results, achieving statistical differences in one of the scenarios, which makes it a good option to tackle this kind of problems.

\notea{
\section*{Acknowledgement}
Javier Fumanal Idocin's, Jose Antonio Sanz's, Javier Fernandez's, and Humberto Bustince's research has been supported by the project PID2019-108392GB I00 (AEI/10.13039/501100011033).
}

\bibliographystyle{unsrt}
\bibliography{bci}

\begin{thebibliography}{10}

\bibitem{lin2018wireless}
Chin-Teng Lin, Ching-Yu Chiu, Avinash~Kumar Singh, Jung-Tai King, Li-Wei Ko,
  Yun-Chen Lu, and Yu-Kai Wang.
\newblock A wireless multifunctional ssvep-based brain--computer interface
  assistive system.
\newblock {\em IEEE Transactions on Cognitive and Developmental Systems},
  11(3):375--383, 2018.

\bibitem{wang2015eeg}
Yu-Kai Wang, Tzyy-Ping Jung, and Chin-Teng Lin.
\newblock Eeg-based attention tracking during distracted driving.
\newblock {\em IEEE transactions on neural systems and rehabilitation
  engineering}, 23(6):1085--1094, 2015.

\bibitem{ko2019multimodal}
Li-Wei {Ko}, Yi-Chen {Lu}, Humberto {Bustince}, Yu-Cheng {Chang}, Yang {Chang},
  Javier {Ferandez}, Yu-Kai {Wang}, Jose~Antonio {Sanz}, Gracaliz~Pereira
  {Dimuro}, and Chin-Teng {Lin}.
\newblock Multimodal fuzzy fusion for enhancing the motor-imagery-based brain
  computer interface.
\newblock {\em IEEE Computational Intelligence Magazine}, 14(1):96--106, 2019.

\bibitem{lance2015towards}
Brent~J Lance, Jon Touryan, Yu-Kai Wang, Shao-Wei Lu, Chun-Hsiang Chuang, Peter
  Khooshabeh, Paul Sajda, Amar Marathe, Tzyy-Ping Jung, Chin-Teng Lin, et~al.
\newblock Towards serious games for improved bci.
\newblock {\em Handbook of Digital Games and Entertainment Technologies.
  Singapore: Springer}, pages 1--28, 2015.

\bibitem{tabar2016novel}
Yousef~Rezaei Tabar and Ugur Halici.
\newblock A novel deep learning approach for classification of eeg motor
  imagery signals.
\newblock {\em Journal of neural engineering}, 14(1):016003, 2016.

\bibitem{8386437}
Y.~{Zhang}, C.~S. {Nam}, G.~{Zhou}, J.~{Jin}, X.~{Wang}, and A.~{Cichocki}.
\newblock Temporally constrained sparse group spatial patterns for motor
  imagery bci.
\newblock {\em IEEE Transactions on Cybernetics}, 49(9):3322--3332, Sep. 2019.

\bibitem{blankertz2011single}
Benjamin Blankertz, Steven Lemm, Matthias Treder, Stefan Haufe, and
  Klaus-Robert M{\"u}ller.
\newblock Single-trial analysis and classification of erp components—a
  tutorial.
\newblock {\em NeuroImage}, 56(2):814--825, 2011.

\bibitem{7300244}
M.~{Mironovova} and J.~{Bíla}.
\newblock Fast fourier transform for feature extraction and neural network for
  classification of electrocardiogram signals.
\newblock In {\em 2015 Fourth International Conference on Future Generation
  Communication Technology (FGCT)}, pages 1--6, 2015.

\bibitem{zheng2019emotions}
W.~{Zheng}, W.~{Liu}, Y.~{Lu}, B.~{Lu}, and A.~{Cichocki}.
\newblock Emotionmeter: A multimodal framework for recognizing human emotions.
\newblock {\em IEEE Transactions on Cybernetics}, 49(3):1110--1122, March 2019.

\bibitem{murugappan2014wireless}
M~Murugappan, Subbulakshmi Murugappan, Celestin Gerard, et~al.
\newblock Wireless eeg signals based neuromarketing system using fast fourier
  transform (fft).
\newblock In {\em 2014 IEEE 10th International Colloquium on Signal Processing
  and its Applications}, pages 25--30. IEEE, 2014.

\bibitem{iacovello2016stimuli}
D.~{Iacoviello}, A.~{Petracca}, M.~{Spezialetti}, and G.~{Placidi}.
\newblock A classification algorithm for electroencephalography signals by
  self-induced emotional stimuli.
\newblock {\em IEEE Transactions on Cybernetics}, 46(12):3171--3180, Dec 2016.

\bibitem{8337847}
L.~{Xie}, Z.~{Deng}, P.~{Xu}, K.~{Choi}, and S.~{Wang}.
\newblock Generalized hidden-mapping transductive transfer learning for
  recognition of epileptic electroencephalogram signals.
\newblock {\em IEEE Transactions on Cybernetics}, 49(6):2200--2214, June 2019.

\bibitem{iran2017tfs}
A.~{Jafarifarmand}, M.~A. {Badamchizadeh}, S.~{Khanmohammadi}, M.~A. {Nazari},
  and B.~M. {Tazehkand}.
\newblock A new self-regulated neuro-fuzzy framework for classification of eeg
  signals in motor imagery bci.
\newblock {\em IEEE Transactions on Fuzzy Systems}, 26(3):1485--1497, June
  2018.

\bibitem{nott2017tfs}
P.~A. {Herman}, G.~{Prasad}, and T.~M. {McGinnity}.
\newblock Designing an interval type-2 fuzzy logic system for handling
  uncertainty effects in brain–computer interface classification of motor
  imagery induced eeg patterns.
\newblock {\em IEEE Transactions on Fuzzy Systems}, 25(1):29--42, Feb 2017.

\bibitem{multi2019tfs}
T.~K. {Reddy}, V.~{Arora}, L.~{Behera}, Y.~{Wang}, and C.~{Lin}.
\newblock Multiclass fuzzy time-delay common spatio-spectral patterns with
  fuzzy information theoretic optimization for eeg-based regression problems in
  brain–computer interface (bci).
\newblock {\em IEEE Transactions on Fuzzy Systems}, 27(10):1943--1951, Oct
  2019.

\bibitem{JIN2019262}
Jing Jin, Yangyang Miao, Ian Daly, Cili Zuo, Dewen Hu, and Andrzej Cichocki.
\newblock Correlation-based channel selection and regularized feature
  optimization for mi-based bci.
\newblock {\em Neural Networks}, 118:262 -- 270, 2019.

\bibitem{feng2018towards}
Jiankui Feng, Erwei Yin, Jing Jin, Rami Saab, Ian Daly, Xingyu Wang, Dewen Hu,
  and Andrzej Cichocki.
\newblock Towards correlation-based time window selection method for motor
  imagery bcis.
\newblock {\em Neural Networks}, 102:87--95, 2018.

\bibitem{qi2020spatiotemporal}
Feifei Qi, Wei Wu, Zhu~Liang Yu, Zhenghui Gu, Zhenfu Wen, Tianyou Yu, and
  Yuanqing Li.
\newblock Spatiotemporal-filtering-based channel selection for single-trial eeg
  classification.
\newblock {\em IEEE Transactions on Cybernetics}, 2020.

\bibitem{sugeno1974theory}
Michio Sugeno.
\newblock Theory of fuzzy integrals and its applications.
\newblock {\em Doct. Thesis, Tokyo Institute of technology}, 1974.

\bibitem{beliakov2016practical}
Gleb Beliakov, Humberto Bustince, and Tomasa~Calvo S{\'a}nchez.
\newblock {\em A practical guide to averaging functions}, volume 329.
\newblock Springer, 2016.

\bibitem{grabisch2000application}
Michel Grabisch and Marc Roubens.
\newblock Application of the choquet integral in multicriteria decision making.
\newblock {\em Fuzzy Measures and Integrals-Theory and Applications}, pages
  348--374, 2000.

\bibitem{dias2018using}
Camila~Alves Dias, J{\'e}ssica~CS Bueno, Eduardo~N Borges, Silvia~SC Botelho,
  Gra{\c{c}}aliz~Pereira Dimuro, Giancarlo Lucca, Javier Fernand{\'e}z,
  Humberto Bustince, and Paulo Lilles Jorge~Drews Junior.
\newblock Using the choquet integral in the pooling layer in deep learning
  networks.
\newblock In {\em North American Fuzzy Information Processing Society Annual
  Conference}, pages 144--154. Springer, 2018.

\bibitem{auephanwiriyakul2002generalized}
Sansanee Auephanwiriyakul, James~M Keller, and Paul~D Gader.
\newblock Generalized choquet fuzzy integral fusion.
\newblock {\em Information fusion}, 3(1):69--85, 2002.

\bibitem{murofushi1991fuzzy}
Toshiaki Murofushi and Michio Sugeno.
\newblock Fuzzy t-conorm integral with respect to fuzzy measures:
  generalization of sugeno integral and choquet integral.
\newblock {\em Fuzzy Sets and Systems}, 42(1):57--71, 1991.

\bibitem{dimuro2020state}
Gra{\c{c}}aliz~Pereira Dimuro, Javier Fern{\'a}ndez, Benjam{\'\i}n Bedregal,
  Radko Mesiar, Jos{\'e}~Antonio Sanz, Giancarlo Lucca, and Humberto Bustince.
\newblock The state-of-art of the generalizations of the choquet integral: From
  aggregation and pre-aggregation to ordered directionally monotone functions.
\newblock {\em Information Fusion}, 57:27--43, 2020.

\bibitem{LUCCA201894}
Giancarlo Lucca, José~Antonio Sanz, Graçaliz~Pereira Dimuro, Benjamín
  Bedregal, Humberto Bustince, and Radko Mesiar.
\newblock Cf-integrals: A new family of pre-aggregation functions with
  application to fuzzy rule-based classification systems.
\newblock {\em Information Sciences}, 435:94 -- 110, 2018.

\bibitem{dimuro2020generalized}
Gra{\c{c}}aliz~Pereira Dimuro, Giancarlo Lucca, Benjam{\'\i}n Bedregal, Radko
  Mesiar, Jos{\'e}~Antonio Sanz, Chin-Teng Lin, and Humberto Bustince.
\newblock Generalized cf1f2-integrals: From choquet-like aggregation to ordered
  directionally monotone functions.
\newblock {\em Fuzzy Sets and Systems}, 378:44--67, 2020.

\bibitem{yager2012ordered}
Ronald~R Yager and Janusz Kacprzyk.
\newblock {\em The ordered weighted averaging operators: theory and
  applications}.
\newblock Springer Science \& Business Media, 2012.

\bibitem{yager2004generalized}
Ronald~R Yager.
\newblock Generalized owa aggregation operators.
\newblock {\em Fuzzy Optimization and Decision Making}, 3(1):93--107, 2004.

\bibitem{de2017algorithm}
Laura De~Miguel, Mikel Sesma-Sara, Mikel Elkano, M~Asiain, and Humberto
  Bustince.
\newblock An algorithm for group decision making using n-dimensional fuzzy
  sets, admissible orders and owa operators.
\newblock {\em Information Fusion}, 37:126--131, 2017.

\bibitem{merigo2011uncertain}
Jose~M Merigo and Montserrat Casanovas.
\newblock The uncertain generalized owa operator and its application to
  financial decision making.
\newblock {\em International Journal of Information Technology \& Decision
  Making}, 10(02):211--230, 2011.

\bibitem{bustince2010overlap}
H~Bustince, J~Fernandez, R~Mesiar, Javier Montero, and R~Orduna.
\newblock Overlap functions.
\newblock {\em Nonlinear Analysis: Theory, Methods \& Applications},
  72(3-4):1488--1499, 2010.

\bibitem{DEMIGUEL201981}
Laura~De Miguel, Daniel Gómez, J.~Tinguaro Rodríguez, Javier Montero,
  Humberto Bustince, Graçaliz~P. Dimuro, and José~Antonio Sanz.
\newblock General overlap functions.
\newblock {\em Fuzzy Sets and Systems}, 372:81 -- 96, 2019.
\newblock Theme: Aggregation Operations.

\bibitem{elkano2016fuzzy}
Mikel Elkano, Mikel Galar, Jose Sanz, and Humberto Bustince.
\newblock Fuzzy rule-based classification systems for multi-class problems
  using binary decomposition strategies: on the influence of n-dimensional
  overlap functions in the fuzzy reasoning method.
\newblock {\em Information Sciences}, 332:94--114, 2016.

\bibitem{elkano2014enhancing}
Mikel Elkano, Mikel Galar, Jose~Antonio Sanz, Alberto Fern{\'a}ndez, Edurne
  Barrenechea, Francisco Herrera, and Humberto Bustince.
\newblock Enhancing multiclass classification in farc-hd fuzzy classifier: On
  the synergy between $ n $-dimensional overlap functions and decomposition
  strategies.
\newblock {\em IEEE Transactions on Fuzzy Systems}, 23(5):1562--1580, 2014.

\bibitem{wu2016fuzzy}
Shang-Lin Wu, Yu-Ting Liu, Tsung-Yu Hsieh, Yang-Yin Lin, Chih-Yu Chen,
  Chun-Hsiang Chuang, and Chin-Teng Lin.
\newblock Fuzzy integral with particle swarm optimization for a
  motor-imagery-based brain--computer interface.
\newblock {\em IEEE Transactions on Fuzzy Systems}, 25(1):21--28, 2016.

\bibitem{tangermann2012review}
Michael Tangermann, Klaus-Robert M{\"u}ller, Ad~Aertsen, Niels Birbaumer,
  Christoph Braun, Clemens Brunner, Robert Leeb, Carsten Mehring, Kai~J Miller,
  Gernot Mueller-Putz, et~al.
\newblock Review of the bci competition iv.
\newblock {\em Frontiers in neuroscience}, 6:55, 2012.

\bibitem{GUPTA1991431}
M.M. Gupta and J.~Qi.
\newblock Theory of t-norms and fuzzy inference methods.
\newblock {\em Fuzzy Sets and Systems}, 40(3):431 -- 450, 1991.

\bibitem{tfspreagregaciones}
Giancarlo Lucca, Jose Antonio~Sanz, Gracaliz Pereira~Dimuro, Benjamin Bedregal,
  Radko Mesiar, Anna Kolesarova, and Humberto Bustince.
\newblock {Preaggregation Functions: Construction and an Application}.
\newblock {\em {IEEE Transactions on Fuzzy Systems}}, {24}({2}):{260--272},
  {APR} {2016}.

\bibitem{akin2002comparison}
Mehmet Akin.
\newblock Comparison of wavelet transform and fft methods in the analysis of
  eeg signals.
\newblock {\em Journal of medical systems}, 26(3):241--247, 2002.

\bibitem{teplan2002fundamentals}
Michal Teplan et~al.
\newblock Fundamentals of eeg measurement.
\newblock {\em Measurement science review}, 2(2):1--11, 2002.

\bibitem{blankertz2007optimizing}
Benjamin Blankertz, Ryota Tomioka, Steven Lemm, Motoaki Kawanabe, and
  Klaus-Robert Muller.
\newblock Optimizing spatial filters for robust eeg single-trial analysis.
\newblock {\em IEEE Signal processing magazine}, 25(1):41--56, 2007.

\bibitem{guger2000real}
Christoph Guger, Herbert Ramoser, and Gert Pfurtscheller.
\newblock Real-time eeg analysis with subject-specific spatial patterns for a
  brain-computer interface (bci).
\newblock {\em IEEE transactions on rehabilitation engineering}, 8(4):447--456,
  2000.

\bibitem{gramfort2013meg}
Alexandre Gramfort, Martin Luessi, Eric Larson, Denis~A Engemann, Daniel
  Strohmeier, Christian Brodbeck, Roman Goj, Mainak Jas, Teon Brooks, Lauri
  Parkkonen, et~al.
\newblock Meg and eeg data analysis with mne-python.
\newblock {\em Frontiers in neuroscience}, 7:267, 2013.

\bibitem{tibarewala2010performance}
S.~{Bhattacharyya}, A.~{Khasnobish}, S.~{Chatterjee}, A.~{Konar}, and D.~N.
  {Tibarewala}.
\newblock Performance analysis of lda, qda and knn algorithms in left-right
  limb movement classification from eeg data.
\newblock In {\em 2010 International Conference on Systems in Medicine and
  Biology}, pages 126--131, Dec 2010.

\bibitem{8858815}
G.~{Lucca}, J.~A. {Sanz}, G.~P. {Dimuro}, E.~N. {Borges}, H.~{Santos}, and
  H.~{Bustince}.
\newblock Analyzing the performance of different fuzzy measures with
  generalizations of the choquet integral in classification problems.
\newblock In {\em 2019 IEEE International Conference on Fuzzy Systems
  (FUZZ-IEEE)}, pages 1--6, 2019.

\bibitem{arroyo1993functional}
Santiago Arroyo, Ronald~P Lesser, Barry Gordon, Sumio Uematsu, Darryl Jackson,
  and Robert Webber.
\newblock Functional significance of the mu rhythm of human cortex: an
  electrophysiologic study with subdural electrodes.
\newblock {\em Electroencephalography and clinical Neurophysiology},
  87(3):76--87, 1993.

\bibitem{wolpaw2002brain}
Jonathan~R Wolpaw, Niels Birbaumer, Dennis~J McFarland, Gert Pfurtscheller, and
  Theresa~M Vaughan.
\newblock Brain--computer interfaces for communication and control.
\newblock {\em Clinical neurophysiology}, 113(6):767--791, 2002.

\bibitem{babiloni1999human}
Claudio Babiloni, Filippo Carducci, Febo Cincotti, Paolo~M Rossini, Christa
  Neuper, Gert Pfurtscheller, and Fabio Babiloni.
\newblock Human movement-related potentials vs desynchronization of eeg alpha
  rhythm: a high-resolution eeg study.
\newblock {\em Neuroimage}, 10(6):658--665, 1999.

\bibitem{pfurtscheller1999event}
Gert Pfurtscheller and FH~Lopes Da~Silva.
\newblock Event-related eeg/meg synchronization and desynchronization: basic
  principles.
\newblock {\em Clinical neurophysiology}, 110(11):1842--1857, 1999.

\bibitem{nguyen2016whole}
Jeffrey {Nguyen}, Frederick {Shipley}, Ashley {Linder}, George {Plummer}, Mochi
  {Liu}, Sagar {Setru}, Joshua {Shaevitz}, and Andrew {Leifer}.
\newblock Whole-brain calcium imaging with cellular resolution in freely
  behaving caenorhabditis elegans.
\newblock {\em Bulletin of the American Physical Society}, 2016.

\bibitem{aguilera2017signatures}
Miguel {Aguilera}, Carlos {Alquézar}, and Eduardo~J. {Izquierdo}.
\newblock Signatures of criticality in a maximum entropy model of the c.
  elegans brain during free behaviour.
\newblock In {\em Proceedings of the 14th European Conference on Artificial
  Life ECAL 2017}, pages 29--35, 2017.

\bibitem{andreas2000topics}
Galka Andreas.
\newblock {\em Topics in nonlinear time series analysis, with implications for
  EEG analysis}, volume~14.
\newblock World Scientific, 2000.

\bibitem{hamilton1994time}
James~D Hamilton.
\newblock {\em Time series analysis}, volume~2.
\newblock Princeton New Jersey, 1994.

\bibitem{cortes1995support}
Corinna Cortes and Vladimir Vapnik.
\newblock Support-vector networks.
\newblock {\em Machine learning}, 20(3):273--297, 1995.

\bibitem{mackay1998introduction}
David~JC MacKay.
\newblock Introduction to gaussian processes.
\newblock {\em NATO ASI Series F Computer and Systems Sciences}, 168:133--166,
  1998.

\bibitem{krausz2003critical}
Gunther Krausz, Reinhold Scherer, G~Korisek, and Gert Pfurtscheller.
\newblock Critical decision-speed and information transfer in the “graz
  brain--computer interface”.
\newblock {\em Applied psychophysiology and biofeedback}, 28(3):233--240, 2003.

\bibitem{vidaurre2007study}
Carmen Vidaurre, Reinhold Scherer, Rafael Cabeza, Alois Schl{\"o}gl, and Gert
  Pfurtscheller.
\newblock Study of discriminant analysis applied to motor imagery bipolar data.
\newblock {\em Medical \& biological engineering \& computing}, 45(1):61, 2007.

\bibitem{kuncheva2003measures}
Ludmila~I Kuncheva and Christopher~J Whitaker.
\newblock Measures of diversity in classifier ensembles and their relationship
  with the ensemble accuracy.
\newblock {\em Machine learning}, 51(2):181--207, 2003.

\bibitem{kuncheva2000independence}
Ludmila~I Kuncheva, Christopher~J Whitaker, Catherine~A Shipp, and Robert~PW
  Duin.
\newblock Is independence good for combining classifiers?
\newblock In {\em Proceedings 15th International Conference on Pattern
  Recognition. ICPR-2000}, volume~2, pages 168--171. IEEE, 2000.

\bibitem{brunner2007spatial}
Clemens Brunner, Muhammad Naeem, Robert Leeb, Bernhard Graimann, and Gert
  Pfurtscheller.
\newblock Spatial filtering and selection of optimized components in four class
  motor imagery eeg data using independent components analysis.
\newblock {\em Pattern recognition letters}, 28(8):957--964, 2007.

\bibitem{mishra2018novel}
Pradeep~Kumar Mishra, B~Jagadish, MPRS Kiran, Pachamuthu Rajalakshmi, and
  D~Santhosh Reddy.
\newblock A novel classification for eeg based four class motor imagery using
  kullback-leibler regularized riemannian manifold.
\newblock In {\em 2018 IEEE 20th International Conference on e-Health
  Networking, Applications and Services (Healthcom)}, pages 1--5. IEEE, 2018.

\bibitem{8452948}
S.~{Selim}, M.~M. {Tantawi}, H.~A. {Shedeed}, and A.~{Badr}.
\newblock A csp\\am-ba-svm approach for motor imagery bci system.
\newblock {\em IEEE Access}, 6:49192--49208, 2018.

\bibitem{RAZI201914}
Sara Razi, Mohammad~Reza {Karami Mollaei}, and Jamal Ghasemi.
\newblock A novel method for classification of bci multi-class motor imagery
  task based on dempster–shafer theory.
\newblock {\em Information Sciences}, 484:14 -- 26, 2019.

\end{thebibliography}

\begin{IEEEbiography}[{\includegraphics[width=1in,height=1.25in,clip,keepaspectratio]{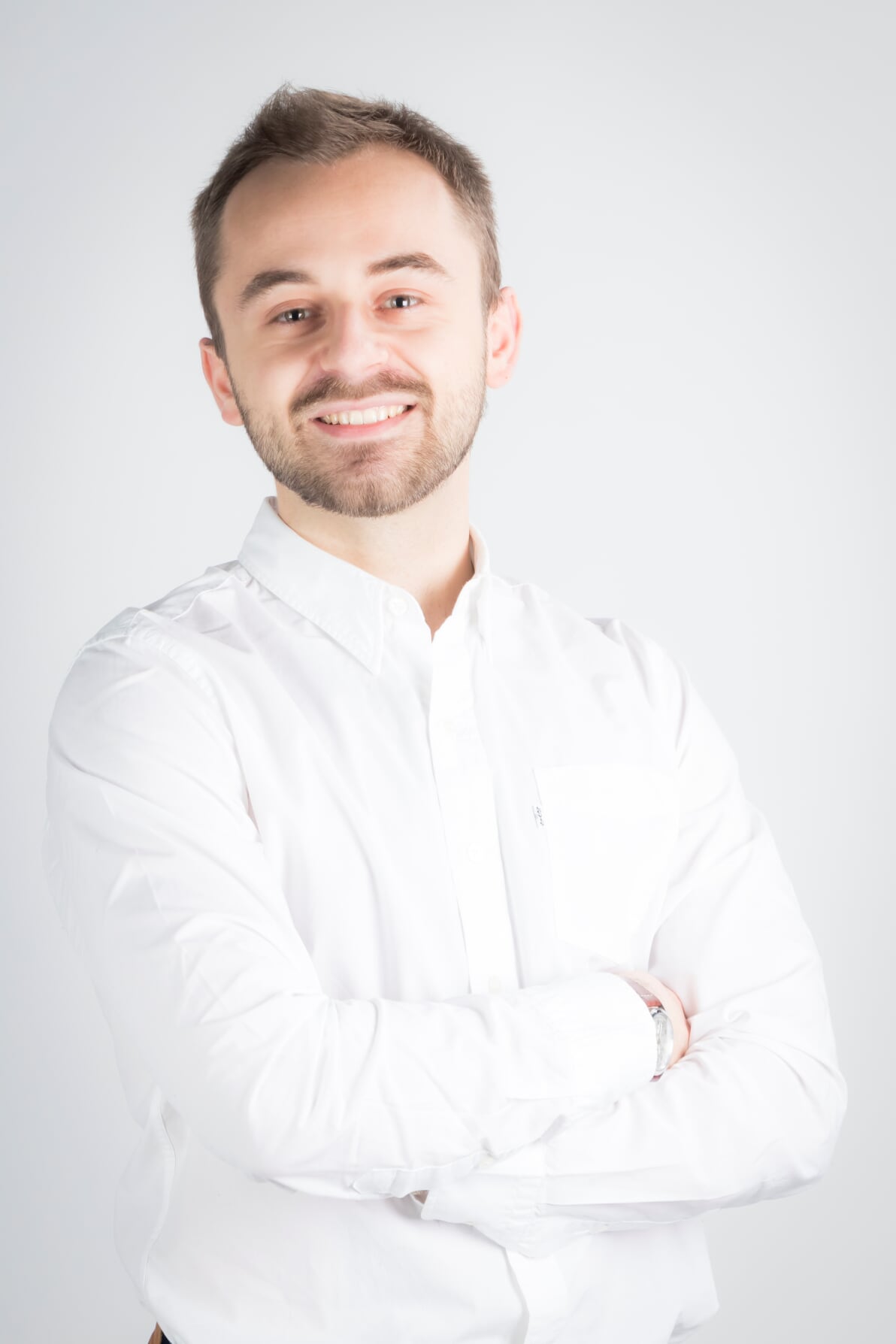}}]{Javier Fumanal Idocin}
	holds a B.Sc in Computer Science at the University of Zaragoza, Spain and a M.Sc in Data Science and Computer Engineering at the University of Granada, Spain. He is now a PhD Student of the Public University of Navarre, Spain in the department of Statistics, Informatics and Mathematics. His research interests include machine intelligence, fuzzy logic, social networks and Brain-Computer Interfaces.
\end{IEEEbiography}

\begin{IEEEbiography}[{\includegraphics[width=1in,height=1.25in,clip,keepaspectratio]{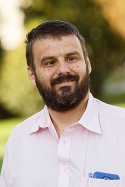}}]{Javier Fernandez}
	received the M.Sc. and Ph.Degrees in Mathematics from the University of Zaragoza, Saragossa, and the Univesity of The Basque Country, Spain, in 1999 and 2003, respectively. He is currently an Associate Lecturer with the with the Department of Statistics, Computer Science and Mathematics Public University of Navarre, Pamplona, Spain. He is the author or coauthor of approximately 45 original articles and is involved with teaching artificial intelligence and computational mathematics for students of the computer sciences and data science. His research interests include fuzzy techniques for image processing, fuzzy
	sets theory, interval-valued fuzzy sets theory, aggregation functions, fuzzy measures, deep learning, stability, evolution equation, and unique continuation.
\end{IEEEbiography}

\begin{IEEEbiography}[{\includegraphics[width=1in,height=1.25in,clip,keepaspectratio]{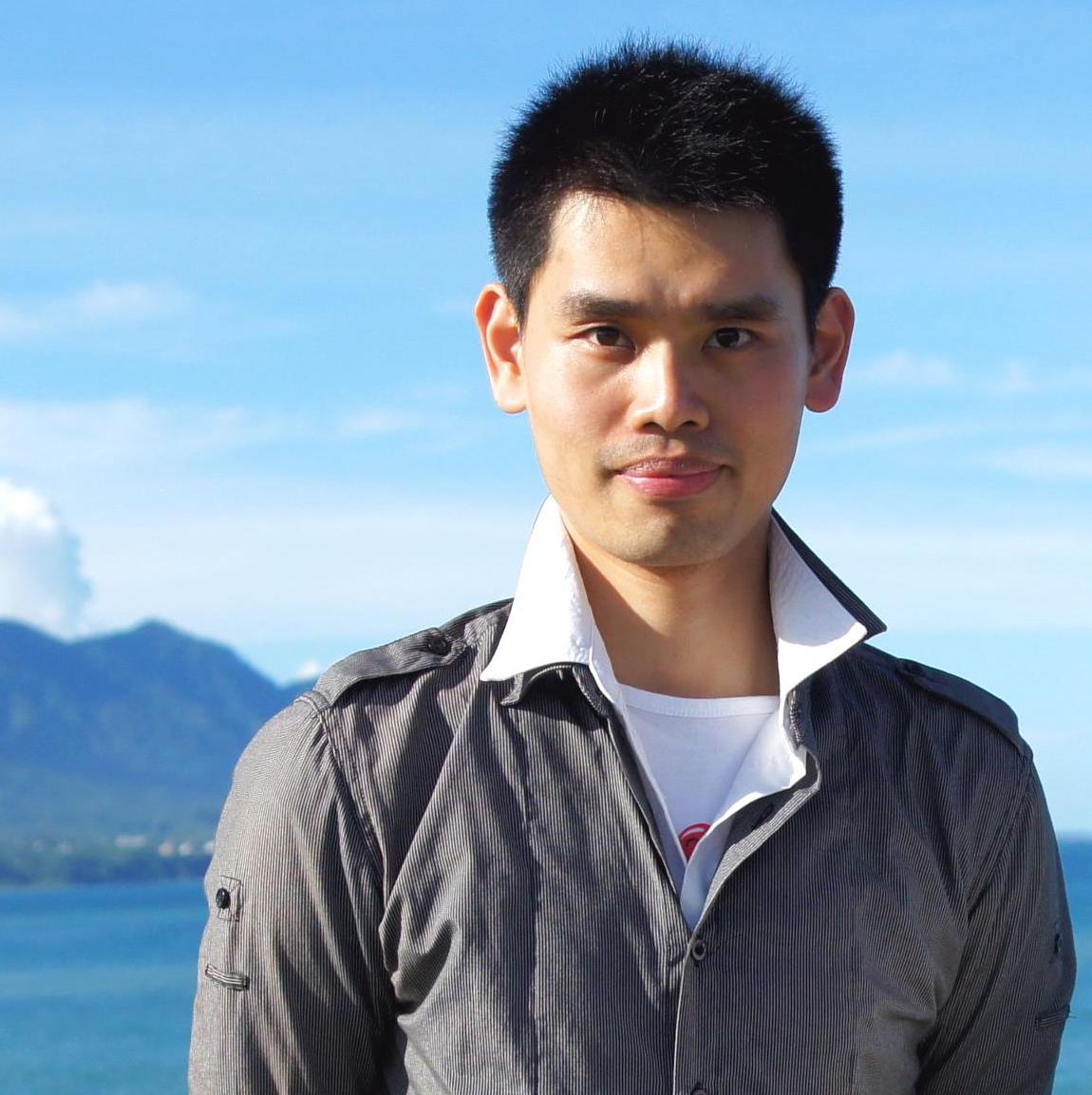}}]{Yu-Kai Wang}
	 (M’13) received the M.S. degree in biomedical engineering and the Ph.D. degree in computer science in 2009 and 2015, respectively, both from the National Chiao Tung University, Taiwan. He is currently a Senior Lecturer of Faculty of Engineering and Information Technology and Co-Director of Motion Platform and Mixed Reality Lab at University of Technology Sydney, Australia. He is the author of 32 published original articles in international journals and more than 40 contributions to international conferences. His current research interests include computational neuroscience, human performance modelling, Brain-Computer Interface and novel human-agent interaction.
\end{IEEEbiography}

\begin{IEEEbiography}[{\includegraphics[width=1in,height=1.25in,clip,keepaspectratio]{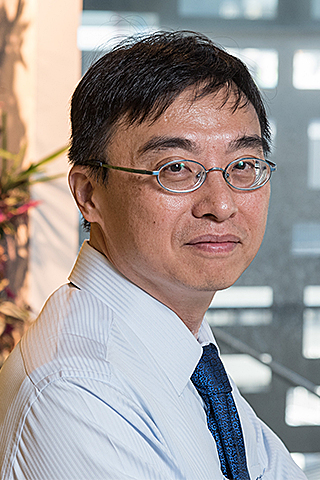}}]{Chin-Teng Lin}
	received the B.S. degree from National Chiao-Tung University (NCTU), Taiwan in 1986, and the Master and Ph.D. degree in electrical engineering from Purdue University, USA in 1989 and 1992, respectively. He is currently the Distinguished Professor of Faculty of Engineering and Information Technology, and Co-Director of Center for Artificial Intelligence, University of Technology Sydney, Australia. He is also invited as Honorary Chair Professor of Electrical and Computer Engineering, NCTU, International Faculty of University of California at San-Diego (UCSD), and Honorary Professorship of University of Nottingham. Dr. Lin was elevated to be an IEEE Fellow for his contributions to biologically inspired information systems in 2005, and was elevated International Fuzzy Systems Association (IFSA) Fellow in 2012. Dr. Lin received the IEEE Fuzzy Systems Pioneer Awards in 2017. He served as the Editor-in-chief of IEEE Transactions on Fuzzy Systems from 2011 to 2016. He also served on the Board of Governors at IEEE Circuits and Systems (CAS) Society in 2005-2008, IEEE Systems, Man, Cybernetics (SMC) Society in 2003-2005, IEEE Computational Intelligence Society in 2008-2010, and Chair of IEEE Taipei Section in 2009-2010. Dr. Lin was the Distinguished Lecturer of IEEE CAS Society from 2003 to 2005 and CIS Society from 2015-2017. He serves as the Chair of IEEE CIS Distinguished Lecturer Program Committee in 2018~2019. He served as the Deputy Editor-in-Chief of IEEE Transactions on Circuits and Systems-II in 2006-2008. Dr. Lin was the Program Chair of IEEE International Conference on Systems, Man, and Cybernetics in 2005 and General Chair of 2011 IEEE International Conference on Fuzzy Systems. Dr. Lin is the coauthor of Neural Fuzzy Systems (Prentice-Hall), and the author of Neural Fuzzy Control Systems with Structure and Parameter Learning (World Scientific). He has published more than 360 journal papers including over 160 IEEE journal papers in the areas of neural networks, fuzzy systems, brain-computer interface, multimedia information processing, cognitive neuro-engineering, and human-machine teaming, that have been cited more than 25,800 times. Currently, his h-index is 72, and his i10-index is 328.
\end{IEEEbiography}

\begin{IEEEbiography}[{\includegraphics[width=1in,height=1.25in,clip,keepaspectratio]{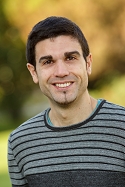}}]{José Antonio Sanz}
	José Antonio Sanz received the M.Sc. and Ph.D. degrees in computer sciences in 2008 and 2011 respectively, both from the Public University of Navarra, Spain. He is currently an assistant professor at the Department of Statistics, Computer Science and Mathematics, Public University of Navarra. He is the author of 36 published original articles in international journals and more than 60 contributions to conferences. He is a member of the European Society for Fuzzy Logic and Technology (EUSFLAT) and the Spanish Association of Artificial Intelligence (AEPIA). He received the best paper award in the FLINS 2012 international conference and the Pepe Millá award in 2014. His research interests include fuzzy techniques for classification problems, interval-valued fuzzy sets, genetic fuzzy systems and medical applications using soft computing techniques.
\end{IEEEbiography}
\begin{IEEEbiography}[{\includegraphics[width=1in,height=1.25in,clip,keepaspectratio]{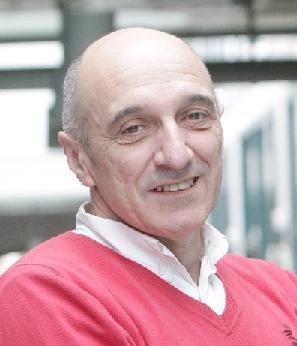}}]{Humberto Bustince}
	received the Graduate degree in physics from the University of Salamanca in 1983 and Ph.D. in mathematics from the Public University of Navarra, Pamplona, Spain, in 1994. He is a Full Professor of Computer Science and Artificial Intelligence in the Public University of Navarra, Pamplona, Spain where he is the main researcher of the Artificial Intelligence and Approximate Reasoning group, whose main research lines are both theoretical (aggregation functions, information and comparison measures, fuzzy sets, and extensions) and applied (image processing, classification, machine learning, data mining, and big data). He has led 11 I+D public-funded research projects, at a national and at a regional level. He is currently the main researcher of a project in the Spanish Science Program and of a scientific network about fuzzy logic and soft computing. He has been in charge of research projects collaborating with private companies. He has taken part in two international research projects. He has authored more than 210 works, according to Web of Science, in conferences and international journals, with around 110 of them in journals of the first quartile of JCR. Moreover, five of these works are also among the highly cited papers of the last ten years, according to Science Essential Indicators of Web of Science. Dr. Bustince is the Editor-in-Chief of the online magazine Mathware \& Soft Computing of the European Society for Fuzzy Logic and technologies and of the Axioms journal. He is an Associated Editor of the IEEE Transactions on Fuzzy Systems Journal and a member of the editorial board of the Journals Fuzzy Sets and Systems, Information Fusion, International Journal of Computational Intelligence Systems and Journal of Intelligent \& Fuzzy Systems. He is the coauthor of a monography about averaging functions and coeditor of several books. He has organized some renowned international conferences such as EUROFUSE 2009 and AGOP. Honorary Professor at the University of Nottingham, National Spanish Computer Science Award in 2019 and EUSFLAT Excellence Research Award in 2019.
\end{IEEEbiography}

\end{document}